\documentclass[a4paper,clock]{article}
\usepackage[dvips]{epsfig}
\usepackage{amssymb}
\usepackage{bm}
\usepackage{dcolumn}
\usepackage{amsmath}

\catcode`\@=11
\def\marginnote#1{}
\newcount\hour
\newcount\minute
\newtoks\amorpm
\hour=\time\divide\hour by60
\minute=\time{\multiply\hour by60 \global\advance\minute
by-\hour}\edef\standardtime{{\ifnum\hour<12
\global\amorpm={am}%
        \else\global\amorpm={pm}\advance\hour by-12 \fi
        \ifnum\hour=0 \hour=12 \fi
        \number\hour:\ifnum\minute<10
0\fi\number\minute\the\amorpm}}
\edef\militarytime{\number\hour:\ifnum\minute<10
0\fi\number\minute}

\def\draftlabel#1{{\@bsphack\if@filesw {\let\thepage\relax
   \xdef\@gtempa{\write\@auxout{\string
      \newlabel{#1}{{\@currentlabel}{\thepage}}}}}\@gtempa
   \if@nobreak \ifvmode\nobreak\fi\fi\fi\@esphack}
        \gdef\@eqnlabel{#1}}
\def\@eqnlabel{}
\def\@vacuum{}
\def\draftmarginnote#1{\marginpar{\raggedright\scriptsize\tt#1}}
\def\draft{\oddsidemargin -.5truein
        \def\@oddfoot{\sl preliminary draft \hfil
        \rm\thepage\hfil\sl\today\quad\militarytime}
        \let\@evenfoot\@oddfoot \overfullrule 3pt
        \let\label=\draftlabel
        \let\marginnote=\draftmarginnote

\def\@eqnnum{(\theequation)\rlap{\kern\marginparsep\tt\@eqnlabel}%
\global\let\@eqnlabel\@vacuum}  }

\def\numberbysection{\@addtoreset{equation}{section}
        \def\theequation{\thesection.\arabic{equation}}}

\def\underline#1{\relax\ifmmode\@@underline#1\else
 $\@@underline{\hbox{#1}}$\relax\fi}

\catcode`@=12
\relax

\numberbysection

\advance \topmargin by -\headheight
\advance \topmargin by -\headsep

\evensidemargin \oddsidemargin
\def\rf#1{(\ref{#1})}
\def\lab#1{\label{#1}}
\def\nn{\nonumber}
\def\br{\begin{eqnarray}}
\def\er{\end{eqnarray}}
\def\be{\begin{equation}}
\def\ee{\end{equation}}

\def\({\left(}
\def\){\right)}

\newcommand{\bi}[1]{\bibitem{#1}}

\relax

\def\a{\alpha}

\def\b{\beta}

\def\d{\delta}
\def\D{\Delta}

\def\g{\gamma}

\def\l{\lambda}
\def\L{\Lambda}

\def\O{\Omega}

\def\pa{\partial}

\def\ra{\rightarrow}
\def\s{\sigma}
\def\S{\Sigma}

\def\tp0{\Theta_{+}^{(0)}}
\def\tm0{\Theta_{-}^{(0)}}

\def\vp{\varphi}

\def\f#1#2#3 {f^{#1#2}_{#3}}

\def\win1{{\sf w_{1+\infty}}}

\def\Win1{{\sf W_{1+\infty}}}

\def\rlx{\relax\leavevmode}
\def\inbar{\vrule height1.5ex width.4pt depth0pt}
\def\IZ{\rlx\hbox{\sf Z\kern-.4em Z}}
\def\IR{\rlx\hbox{\rm I\kern-.18em R}}
\def\IC{\rlx\hbox{\,$\inbar\kern-.3em{\rm C}$}}
\def\IN{\rlx\hbox{\rm I\kern-.18em N}}
\def\IO{\rlx\hbox{\,$\inbar\kern-.3em{\rm O}$}}
\def\IP{\rlx\hbox{\rm I\kern-.18em P}}
\def\IQ{\rlx\hbox{\,$\inbar\kern-.3em{\rm Q}$}}
\def\IF{\rlx\hbox{\rm I\kern-.18em F}}
\def\IG{\rlx\hbox{\,$\inbar\kern-.3em{\rm G}$}}
\def\IH{\rlx\hbox{\rm I\kern-.18em H}}
\def\II{\rlx\hbox{\rm I\kern-.18em I}}
\def\IK{\rlx\hbox{\rm I\kern-.18em K}}
\def\IL{\rlx\hbox{\rm I\kern-.18em L}}
\def\one{\hbox{{1}\kern-.25em\hbox{l}}}
\def\0#1{\relax\ifmmode\mathaccent"7017{#1}%
B        \else\accent23#1\relax\fi}

\def\PRL#1#2#3{{\sl Phys. Rev. Lett.} {\bf#1} (#2) #3}
\def\NPB#1#2#3{{\sl Nucl. Phys.} {\bf B#1} (#2) #3}

\def\PRD#1#2#3{{\sl Phys. Rev.} {\bf D#1} (#2) #3}
\def\PRE#1#2#3{{\sl Phys. Rev.} {\bf E#1} (#2) #3}
\def\PRA#1#2#3{{\sl Phys. Rev.} {\bf A#1} (#2) #3}

\def\PLB#1#2#3{{\sl Phys. Lett.} {\bf #1B} (#2) #3}

\def\AoP#1#2#3{{\sl Ann. of Phys.} {\bf #1} (#2) #3}

\def\IJMPA#1#2#3{{\sl Int. J. Mod. Phys.} {\bf A#1} (#2) #3}

\def\JPAMT#1#2#3{{\sl J. Physics A: Math. Theor.} {\bf A#1} (#2) #3}

\def\JHEP#1#2#3{{\sl JHEP} {\bf #1} (#2) #3}

\def\NJP#1#2#3{{\sl New J. Phys.} {\bf #1} (#2) #3}
\def\UN#1#2#3{{\sl Universe} {\bf #1} (#2) #3}
\def\EPJC#1#2#3{{\sl Eur. Phys. J. C} {\bf #1} (#2) #3}
\def\JLTP #1#2#3{{\sl J Low Temp Phys} {\bf #1} (#2) #3} 
\begin{document}
\begin{titlepage}

\vskip .6in

\begin{center}
{\large {\bf Fermionic back-reaction on kink and topological charge pumping in the $sl(2)$ affine Toda coupled to matter}}
\end{center}

\normalsize
\vskip .4in

\begin{center}

H. Blas$^{a}$ and R. Quica\~no$^{b}$

\par \vskip .2in \noindent

$^{a}$ Instituto de F\'{\i}sica\\
Universidade Federal de Mato Grosso\\
Av. Fernando Correa, $N^{0}$ \, 2367\\
Bairro Boa Esperan\c ca, Cep 78060-900, Cuiab\'a - MT - Brazil. \\
              $^{c}$ Facultad de Ciencias\\ Instituto de Matem\'atica y Ciencias Afines (IMCA)\\
Universidad Nacional de Ingenier\'ia, Av. Tupac Amaru, s/n, Lima-Per\'u.\\
               
\normalsize
\end{center}
\par \vskip .3in \noindent

We explore  the Faddeev-Jackiw (F-J) symplectic Hamiltonian reduction of the $sl(2)$ affine Toda model coupled to matter (ATM), which includes new parametrizations for a scalar field and a Grassmannian fermionic field. The structure of constraints and symplectic potentials primarily dictates the strong-weak dual coupling sectors of the theory, ensuring the equivalence between the Noether and topological currents. The analytical calculations encompass the fermion-kink classical solution, the excited fermion bound states localized on the kink, and the scattering states, all of which account   for the fermion back-reaction on the soliton. The total energy, which includes the classical fermion-soliton interaction energy, the bound-state fermion energy, and the fermion vacuum polarization energy (VPE), is determined by the topological charge of the kink. This system satisfies first-order differential equations and a chiral current conservation equation. Our results demonstrate that the excited fermion bound states and scattering states significantly alter the properties of the kink. Notably, they give rise to a pumping mechanism for the topological charge of the in-gap kink due to fermionic back-reaction, as well as the appearance of kink states in the continuum (KIC).

\end{titlepage}

\section{Introduction}

Integrable models play a pivotal role in theoretical physics, providing insights into the complex dynamics of classical and quantum systems \cite{raja, babelon, frishman}. Among these, the $sl(n)$ affine Toda models coupled to matter (ATM) offer a compelling framework for exploring the interplay between bosonic and fermionic fields. By extending the traditional Toda model to include matter fields, the ATM model captures a broader range of physical phenomena, including nonlinearity, topological defects, chiral confinement, bound states, and correspondence between Noether and topological charges.  These models are known for their ability to describe soliton-fermion configurations and exhibit remarkable properties, making them an invaluable tool for understanding non-linear interactions and topological phenomena \cite{matter, npb1, npb2, prd}.

The back-reaction of fermions on kinks is an area of active research with significant implications for understanding non-perturbative effects in quantum field theories. Kink-fermion systems typically exhibit a fermion zero mode and charge fractionalization \cite{jackiw1}. Additional higher energy valence levels, as excitations of the bound states, can emerge for some models. Recently, kink configurations have been constructed mainly using numerical techniques that account for the back-reaction from the excited fermion bound states \cite{shnir1, shnir2, gani}. The total energy of the fermion-kink system  comprises three components: the classical fermion-soliton interaction energy, the energy of the bound-state fermions, and the fermion vacuum polarization energy (VPE). The VPE, which originates from the interaction of the kink with the Dirac sea, is essential for maintaining the consistency of the semi-classical expansion in the fermion sector and to understand how fermionic back-reaction can modify the stability and dynamics of kinks \cite{weigel1, weigel2}.

An important aspect of our analysis is the examination of the constraint structure and symplectic potentials within the $sl(2)$ ATM model through the Faddeev-Jackiw (F-J) symplectic Hamiltonian reduction \cite{fj, jack1}. These elements determine the nature of the strong-weak dual coupling sectors, providing a framework to ensure the equivalence of Noether and topological currents. Our findings reveal the emergence of fermion excited bound states localized on the kinks of the model. These states are not merely passive features; they actively participate in the dynamics by contributing to back-reaction effects, thereby altering the topological landscape of the system.

Because obtaining exact analytical results for general models is challenging, we use an integrable model to study the effects of fermion back-reaction. To find analytical solutions for this model, we apply tau function techniques, which allow us to construct self-consistent kink-fermion solutions. These techniques allow us to analyze how the kink and fermionic bound states and scattering states properties depend on various model parameters, offering insights into the stability and behavior of these solutions. Our results indicate that the back-reaction of localized and scattering fermions significantly modifies the topological properties of the system. Notably, we observe a topological charge pumping mechanism driven by the fermionic back-reaction, which alters the kink's topological charge and sheds light on the intricate relationship between topology and dynamics in integrable systems.

In contrast to the fermion-soliton models commonly studied in the literature, where the topological charge of the kink is predetermined and associated with degenerate vacua of a self-coupling potential in the scalar field sector, our model allows the asymptotic behavior of the scalar field and the relevant topological charge to be generated dynamically as solutions to a system of first-order equations. An analogous model, in which quantum effects can stabilize a soliton, has been  discussed in \cite{farhi, farhi2}. Our model can be viewed as a specific reduction of that model by setting its scalar self-coupling potential to zero. Indeed, our solitons are classical solutions of the model in \cite{farhi, farhi2} in some regions of parameter space.  

The system of equations of motion is reduced to a set of first-order differential equations as follows.  We reduced the order of the chiral current conservation equation  by introducing a massless free field, $\Sigma$. In this framework, the trivial solution $\Sigma =0$ leads to the equivalence of the Noether and topological currents. Our method differs from the Bogomolnyi trick, which obtains first-order equations by completing the square in the energy functional. However, it is similar to the BPS method in that it expresses soliton energies in terms of topological charges. It also parallels the approach proposed in \cite{devega}, where first-order equations for vortices in 2+1 dimensions were derived by considering the conservation of the energy-momentum tensor. Our analysis is {\sl quasiclassical} \cite{weigel1, weigel2, lee}, but investigating the impact of quantum corrections would be an intriguing direction for future research. Notably, we have a kink-fermion configuration energy and a fermion bound state energy, both of which are lower than the energy of a single free fermion.
 
In the analysis of quantum effects in kink solitons coupled to a single excited fermion bound state within a semi-classical framework, equal importance must be given to the energy of the bound state and the energy of the Dirac sea \cite{weigel1, weigel2}. In the present paper, the Dirac sea energy is computed as the fermion vacuum polarization energy (VPE) and is given equal consideration alongside the fermion-soliton interaction energy and the bound-state fermion energy.

The paper is organized as follows. In section \ref{sec:model} we present the model  and its main symmetries. 
In section \ref{sec:FJ} the F-J reduction process is performed. In sec.  \ref{sec:dual} the gauge fixing and dual sectors are examined in parameter space. In sec. \ref{sec:chi} the chiral confinement and the first order differential equations are discussed. In sec. \ref{sec:bs} the soliton-fermion configurations and spinor bound states are derived. The zero-modes and the excited fermion bound states and the dual sectors are discussed.  In sec. \ref{sec:energy} the energy of kink-fermion plus spinor bound state configurations are computed. In section \ref{sec:diracvacuum} the Dirac sea modification due to the soliton is examined and the total energy is computed. The sec. \ref{sec:discuss} presents the discussions and conclusions. The appendix \ref{app:FJ} presents a brief review of the Faddeev-Jackiw symplectic formalism. The appendix \ref{app1} shows that the first order differential equations imply the second order equation for the scalar field.

\section{The model}
\label{sec:model}
We consider the field theory in $1+1$ dimensions defined by the Lagrangian\footnote{Our notation: $x_{\pm} = t\pm x $, and so, $\pa_{\pm}=\frac{1}{2} (\pa_t \pm \pa_x)$, and $\pa^2=\pa^2_t -\pa^2_x = 4\pa_{-}\pa_{+}$. We use $\g_0 = \(\begin{array}{cc} 0  & i\\
-i & 0\end{array}\)$, $\g_1 = \(\begin{array}{cc} 0  & -i\\
-i & 0\end{array}\)$, $\g_5 = \g_0 \g_1 = \(\begin{array}{cc} 1  & 0\\
0 & -1\end{array}\)$,  and $\psi = \(\begin{array}{c} \psi_{R}\\
\psi_{L}\end{array}\),\,\, \bar{\psi} = \psi^{\dagger} \g_0,\,  \psi_{R} \equiv (\frac{1+\g_5}{2})\psi,\, \psi_{L} \equiv (\frac{1-\g_5}{2})\psi $.} 
\br
\lab{atm0}
{\cal L} =\frac{1}{2}\partial_{\mu }\varphi \partial ^{\mu }\varphi +i\overline{\psi }\gamma ^{\mu}\partial _{\mu }\psi - M \overline{\psi }e^{2 i \hat{\b} \varphi
\gamma _{5}}\psi, 
\er
where $\vp$ is a real scalar field, $\psi$ is a Dirac spinor, $M$ is a mass parameter and $\hat{\b}$ is the coupling constant. This is the so-called  $sl(2)$ affine Toda system coupled to matter field (ATM) \cite{matter, npb2}. Its integrability properties, construction of the general solution including the solitonic ones, the soliton-fermion duality, as well as its symplectic structures were discussed in \cite{npb1, npb2, aop}.  This model has been shown to describe the low-energy effective Lagrangian of QCD$_2$ with one flavor and $N$ colors \cite{prd} and the BCS coupling in spinless fermions in a two dimensional model of high T superconductivity in which the solitons play the role of the Cooper pairs \cite{ferraz}. This model has been earlier studied as a model for fermion confinement in a chiral invariant theory \cite{chang} and the mechanism of fermion mass generation without spontaneously chiral symmetry breaking in two-dimensions \cite{witten, kogut}.

In this paper we will discuss on some special features of the model at the quasi-classical level, as well as new soliton solutions associated  to a Hamiltonian reduced version of the model. The Lagrangian (\ref{atm0}) is invariant under the commuting $U(1)_L \otimes U(1)_R$
left and right local gauge transformations \cite{npb1, npb2}
\br
\lab{leri1}
\vp \ra \vp + \frac{1}{\hat{\b}} \big[\xi_{+}\(x_{+}\) + \xi_{-}\(x_{-}\) \big]\; , 
\er
and
\br
\lab{leri2} 
\psi \ra e^{- i\( 1+ \gamma_5\) \frac{\xi_{+}\(x_{+}\)}{\hat{\b}} 
+ i\( 1- \gamma_5\) \frac{\xi_{-}\(x_{-}\)}{\hat{\b}}}\, \psi 
\; ; \qquad 
{\psi^{*}} \ra e^{ i\( 1+ \gamma_5\) \frac{\xi_{+}\(x_{+}\)}{\hat{\b}} 
- i\( 1- \gamma_5\) \frac{\xi_{-}\(x_{-}\)}{\hat{\b}}} {\psi^{*}}. 
\er
Associated to the global $U(1)$ and $U_5(1)$ transformations one has  
the next currents and conservation laws
\br
J^{\mu} &=& {\bar{\psi}}\, \gamma^{\mu}\, \psi \, , \,\,\, \qquad \qquad \qquad  \qquad
\pa_{\mu}\, J^{\mu} = 0,
\lab{noethersl2}
\\ 
J_5^{\mu} &=&  \bar \psi \gamma^\mu \gamma_5 \psi 
+\frac{1}{\hat{\b}} \partial^\mu \vp  
\; ; \qquad \qquad \pa_{\mu}J_5^{\mu} =0. 
\lab{chiral0}
\er
 
An important feature of the model is the classical equivalence between the $U(1)$ Noether current (\ref{noethersl2}) and the topological current, i.e.
\br
\label{equi11}
 {\bar{\psi}}\, \gamma^{\mu}\, \psi =  \frac{1}{\hat{\b}} \epsilon^{\mu\nu} \pa_{\nu} \, \vp.
\er 
This equivalence holds true for the classical soliton and the zero-mode bound state solutions \cite{npb1}. 

Moreover, the structure of the vacuum of the model (\ref{atm0}) is more complex. The previous literature considered mainly the vacua defined as
\br
\label{vac1}
 \vp_{vac} =  \frac{\pi n}{\hat{\b}},\,\,\, \psi_{vac} =0,\,\,\, \,\,\,\,\ n \in \IZ.  
\er

\section{Faddeev-Jackiw (F-J) reduction and new field parametrizations}
\label{sec:FJ}

We revisit the ATM model and apply the Faddeev-Jackiw symplectic formulation by including new parametrizations for a scalar field and a Grassmannian fermionic field, which allows one to study the intermediate coupling strengths between  the scalar and spinor field configurations, as well as the known strong coupling sine-Gordon and weak coupling massive Thirring sectors of the model. So, let us introduce the next parametrization of the fermion field $\psi$ 
\br
\label{rchi}
 \psi_{R}^{} = \chi_{R}^{}e_{}^{ir\theta}~~ ~~\psi_{L}^{} = \chi_{L}^{}e_{}^{-i r \theta},\,\, \, r=constant,
\er
where $ \chi_{R, L}$ are Grassmannian spinor fields and $\theta$ is a new real scalar field.
So, one has the Lagrangian
 
\be
\lab{lagrangian1}
{\cal L}=\frac{1}{2}\partial_{\mu }\varphi \partial ^{\mu }\varphi +i\overline{\chi }\gamma ^{\mu}\partial _{\mu }\chi + r j^\mu \epsilon_{\mu \nu} \pa^\nu \theta -M \overline{\chi }e^{ 2 i (\hat{\b} \varphi + r \theta) 
\gamma _{5}}\chi +\lambda_{\mu}(2\, \overline{\chi}\gamma^{\mu}\chi- \kappa \,\epsilon^{\mu\nu}\partial_{\nu}(\hat{\b} \varphi + v \theta)),
\ee
where $j^\mu = \bar{\chi}\g^\mu \chi$ and $\kappa = \frac{2}{\hat{\b}^2}$. We have incorporated a gauge fixing term making use of the Lagrange multiplier $\lambda_{\mu}$. The parameter $v$ is a real number, which will be fixed below by requiring a Lorentz invariant reduced Lagrangian. The term incorporating $\l_{\mu}$ in \rf{lagrangian1} will break the left-right local symmetries \rf{leri1}-\rf{leri2} of the ATM model \rf{atm0}. Note that a particular gauge-fixing with $v=0$ in the case $r=0$ of (\ref{rchi})-(\ref{lagrangian1}) has been considered in \cite{aop}.     

Next, we perform the Faddeev-Jackiw symplectic reduction of the model (see appendix \ref{app:FJ} for a brief review of this formalism). So, in order to write \rf{lagrangian1} in the first order form as required  by \rf{lag1} let us rewrite  it as
\br
\nn
 {\cal L}
 & =  & \frac{1}{2}(\dot{\varphi}^{2}_{} - {\varphi'}^{2}_{}) 
 + i \chi^{*}_{L}\dot{\chi}_{L}^{} + i \chi^{*}_{R}\dot{\chi}_{R}^{} +  i \chi^{*}_{L} \chi_{L}^{'} - i \chi^{*}_{R} \chi_{R}^{'}  + r \dot{\theta} j^{1}_{} + r \theta' j^{0}_{} + \\ 
\nn
 &&  i M e^{2i(\hat{\b}\varphi + r\theta)}_{} \chi^{*}_{L} \chi^{}_{R} - i M e^{-2i(\hat{\b} \varphi + r\theta)}_{} \chi^{*}_{R} \chi^{}_{L} + \\
&&\lambda_{0}^{}( 2 j^{0}_{} - \kappa \hat{\b} \varphi' - \kappa v \theta') + 
  \lambda_{1}^{}( 2 j^{1} + \kappa \hat{\b} \dot{\varphi} + \kappa v \dot{\theta}). \label{lag2}
\er
Next, let us calculate the conjugated momenta 
\br
\lab{moments}
\nn
\pi_{R}^{} &=&  -i\chi_{R}^{*}, \,\,\,\,\,\,\,\,\,\,\,\, \pi_{L}^{} =   -i\chi_{L}^{*},\,\,\,\,\,
  \pi _{\lambda_{\mu}} = 0,\,\,\,\,\,\,\,   \pi^{}_{\theta} = \kappa  v \lambda^{}_{1} + r j^{1}_{},\,\,\,\,\,
 \pi_{\varphi} = \dot{\varphi}+ \kappa \hat{\b}\lambda_{1}.
\er
We are assuming the Dirac fields as anti-commuting Grasmannian variables and their momenta variables defined through {\bf left} derivatives.   
Then, as usual, the Hamiltonian is defined by
\be
\lab{hamiltonian}
{\cal H}_{c}=\dot{\varphi}\pi _{\varphi} + \dot{\theta}\pi _{\theta} +  \dot{\chi}_{R}^{}\pi_{R}^{} + \dot{\chi}_{L}^{}\pi _{L}^{} - \cal{L}.
\ee
Then, the Hamiltonian density becomes 
\br
\lab{hamiltonian1}
\nn
{\cal H}_{c}^{} & = & \frac{1}{2} \pi _{\varphi}^{2}+ \frac{1}{2} (\hat{\b} \kappa)^2 \lambda_{1}^2+\frac{1}{2}{\varphi'}_{}^{2}  + \pi_{L}^{}\chi_{L}^{\prime} - \pi_{R}^{}\chi_{R}^{\prime} -
2\lambda_{1}^{} j_{}^{1} -  \kappa \hat{\b} \lambda_{1}^{}\pi_{\varphi}^{} - \\
&& \lambda_{0}^{}(2j^{0}_ {} - \kappa \hat{\b}\varphi^{\prime}_{} - \kappa  v\theta') - r\theta'j^{0}_{} +  i M (e^{-2i(\hat{\b} \varphi + r\theta)}_{} \chi^{*}_{R}\chi^{}_{L} - e^{2i(\hat{\b} \varphi + r\theta)}_{} \chi^{*}_{L}\chi^{}_{R}). 
\er
Now, the same Legendre transform \rf{hamiltonian} is used to write the first order Lagrangian 
\be
 \lab{lagran3}
 {\cal L} = \dot{\varphi}\pi _{\varphi} + \dot{\theta}\pi^{}_{\theta}  + \dot{\chi}_{R}^{}\pi_{R}^{} + \dot{\chi}_{L}^{}\pi _{L}^{} - {\cal H}_{c}.
\ee
Our starting point for the F-J analysis will be this first order Lagrangian. The Lagrangian \rf{lagran3} is already in the form \rf{lag1}, and the Euler-Lagrange equations for the components of the Lagrange multiplier $\l_{\mu}$ allow us to solve one of them 
\br
\label{la1}
\lambda^{}_{1} = \frac{2}{(\kappa \hat{\b})^2} j_{}^{1}+ \frac{1}{\kappa \hat{\b}} \pi_{\varphi}^{}\, ,
\er
and the $\l_0$ component leads to the constraint
\be
\lab{cons}
\O_{1}^{} \, \equiv \, 2j_{}^{0} - \kappa \hat{\b} \varphi' - \kappa  v\theta' = 0\, .
\ee
The constraint \rf{cons} can be solved as
\br
\label{constsol}
 \varphi = - \frac{v}{\hat{\b}}\theta + \hat{\b} \int^{x}_{-\infty} dx \, j^{0}.
\er
Making use of the conservation law $\pa_{\mu} (\bar{\chi} \g^{\mu} \chi) =0$, which is inherited from (\ref{noethersl2}) by the spinor $\chi$, the expression (\ref{constsol}) can be written as
\br
\label{vpt}
 \dot{\varphi} =  -\frac{v}{\hat{\b}} \dot{\theta} - \hat{\b} j^{1}_{}.
\er  
Next, the Lagrange multiplier $\lambda_{1}$ in \rf{la1} and the field $\vp$ in the form (\ref{constsol}) must be replaced back into the Hamiltonian (\ref{hamiltonian1}). Moreover,  the time-derivative $\dot{\vp}$ expression in (\ref{vpt}) must be replaced into the first term of the Lagrangian (\ref{lagran3}). Thus, we get the following Lagrangian 
\begin{eqnarray}
\nonumber
{\cal L}^{\prime}& = &  \dot{\theta}\pi^{}_{\theta} - \frac{v}{\hat{\b}}\dot{\theta}\pi_{\varphi}^{}   - \frac{1}{2} (\frac{v}{\hat{\b}})^2{\theta'}_{}^{2}  +(v+r)\theta' j^{0}_{} + \frac{1}{2} \hat{\b}^2
        (j^{1}_{})^{2} - \frac{1}{2} \hat{\b}^2 (j^{0}_{})^{2} +         
 \, i\chi_{R}^{*}\dot{\chi}_{R}^{} + i\chi_{L}^{*}\dot{\chi}_{L}^{}
	-\, i\chi_{R}^{*}\chi_{R}^{\prime} + i\chi_{L}^{*}\chi_{L}^{\prime} + \\
& &   \, i Me^{2i((r - v)\theta +  \hat{\b}^2\int^{x}dx j^{0})}_{}\chi^{*}_{L}\chi_{R}^{}
      -\, i Me^{-2i((r - v)\theta + \hat{\b}^2 \int^{x} dx j^{0})}_{}\chi^{*}_{R}\chi_{L}^{}.  \label{lagran2}
\end{eqnarray}

In order to covariantize the Lagrangian (\ref{lagran2}) let us assume
\br
\label{vbeta}
v = \hat{\b},
\er  
supplied with the following transformations 
\begin{eqnarray}
 \lab{transf1}
 \nn
 \pi_{\varphi}^{} & \mapsto & -(\hat{\b} + r) j_{}^{1},\,\,\,\,\,\,\,\,\,\,\,\,
 \pi_{\theta}^{}   \mapsto  \frac{1}{2} \dot{\theta}\\
 \chi_{R}         & \mapsto & e^{- i \hat{\b}^2  \int^{x}_{-\infty}dx\,j^{0}}_{}\xi_{R}^{},\,\,\,\,\,\,\,
 \chi_{L}^{}       \mapsto  e^{i \hat{\b}^2 \int^{x}_{-\infty}dx\,j^{0}}_{}\xi_{L}^{}.
\end{eqnarray}
So, one has the following covariant Lagrangian
\begin{eqnarray}
\nonumber
 {\cal L} & = & \frac{1}{2} \partial^{\mu}_{}\theta\partial^{}_{\mu}\theta  +
                                                            (r+\hat{\b})j^{\mu}_{}\epsilon^{}_{\mu\nu}\partial^{\nu}_{}\theta - \frac{3}{2} \hat{\b}^2  j^{\mu}_{}j^{}_{\mu} +
                                                            i\overline{\xi}\gamma^{\mu}_{}\partial^{}_{\mu}\xi \\
&& 
                                                        - M \, \overline{\xi}\xi\cos[2(r-\hat{\b})\theta] +
                                                            i M \,  \overline{\xi}\gamma^{5}_{}\xi\sin[2(r-\hat{\b})\theta].  \label{atmfj00}
\end{eqnarray} 
Next, let us define 
\br
\nonumber
\a &\equiv& r + \hat{\b},\\
\label{abg}
\b &\equiv& 2 (r - \hat{\b}),\\
\nonumber
g &\equiv&  \frac{3}{2} \hat{\b}^2\, \rightarrow  g = \frac{3}{32} (2 \a - \b)^2. 
\er
Then, we are left with the Lagrangian
\br
\nonumber
{\cal L} & = &\frac{1}{2} \partial^{\mu}_{}\theta\partial^{}_{\mu}\theta + \alpha
                                                             j^{\mu}_{}\epsilon^{}_{\mu\nu}\partial^{\nu}_{}\theta - g  j^{\mu}_{}j^{}_{\mu} +
                                                            i\overline{\xi}\gamma^{\mu}_{}\partial^{}_{\mu}\xi \\
&&
                                                       - M  \, \overline{\xi}\xi\cos{(\beta \theta)} +
                                                            i M   \, \overline{\xi}\gamma^{5}_{}\xi\sin{(\beta \theta)}. \lab{mthsg} 
\er 
This reduced Lagrangian is defined for the real scalar field $\theta$ and the Dirac spinor $\xi$, and the three independent parameters $\{M, \a, \b\}$. The equations of motion following from the Lagrangian \rf{mthsg} become
\br
\label{thetaeq}
\pa^2 \theta - \alpha \epsilon^{\mu \nu} \pa_{\mu} j_{\nu} - \beta\, M   \bar{\xi}\xi \sin{(\beta \theta)} - i \beta\, M  \bar{\xi}\gamma_5\xi \cos{(\beta \theta)} &=&0\\
\label{psieq}
i\gamma^{\mu} \pa_{\mu} \xi + 2g \gamma^{\mu} \xi \( \frac{\alpha}{2g} \epsilon_{\mu \nu} \pa^{\nu} \theta -   j_{\mu}\) - M   \xi \cos{(\beta \theta)} +  i M  \gamma_5\xi \sin{(\beta \theta)} &=&0.\\
\label{bpsieq}
-i \pa_{\mu} \bar{\xi} \gamma^{\mu} + 2 g \( \frac{\alpha}{2g} \epsilon_{\mu \nu} \pa^{\nu} \theta -   j_{\mu}\) \bar{\xi} \gamma^{\mu}  - M   \bar{\xi} \cos{(\beta \theta)} +  i M   \bar{\xi}\gamma_5 \sin{(\beta \theta)} &=&0. 
\er
The Lagrangian \rf{mthsg} possesses the next global symmetries:
the $U(1):$ $\xi \rightarrow e^{i \delta} \xi,\,\,\,(\delta =const.)$  and $U(1)$ chiral: $\xi \rightarrow e^{i \zeta \g_5} \xi,\,\, \theta \rightarrow \theta + \frac{2 \zeta}{\beta},\,\,\,(\zeta =const.)$ symmetries, respectively. The associated Noether currents and conservation laws become, respectively
\br
\label{vector}
j^{\mu}    \equiv  \bar{\xi}\gamma^{\mu} \xi\qquad \qquad \qquad \qquad \,\,\,\,\,\,\,\,\,\,&\Rightarrow& \pa_{\mu} j^{\mu} =0;\\
\label{chiral}
j_5^{\mu} \equiv  - \pa^{\mu} \theta + (\alpha+\frac{\beta}{2})\, \bar{\xi} \gamma^{\mu} \gamma_5\xi\,\,\,\,\,\,\,\,\, &\Rightarrow & \pa_{\mu} j_5^{\mu} =0.
\er  
 
For the special value $\a = 0$, the model  (\ref{mthsg}) exhibits the left/right local symmetry of the type  (\ref{leri1})-(\ref{leri2}). Therefore, for this particular parameter value one must impose a gauge fixing condition to the Lagrangian (\ref{mthsg}). This gauge fixing procedure will be performed below in order to inspect the strong coupling sector of the model.  

\section{Parameter space, gauge fixing  and dual sectors}
\label{sec:dual}
In this section we will examine  the model (\ref{mthsg}) by choosing some particular values for the set of parameters $\{\a, \b\}$. This process will reproduce either the weak coupling spinor or the  strong coupling scalar sector of the model. So, this procedure will provide the massive Thirring model plus massless free scalar, as well as the sine-Gordon model.

\subsection{Massive Thirring model plus massless free scalar field}
 \label{sec:MTS}
Let us consider the case $r= \hat{\beta} \rightarrow \beta=0$. This choice of parameters make the fields $\cos{(\b \theta)}$ and $\sin{(\b \theta)}$ in the second line of the Lagrangian (\ref{mthsg}) to be constants. Moreover, the decoupling procedure of the spinor from the scalar $\theta$ will be performed below. In this case one has $ g= \frac{3}{8}\alpha^2$ and the Lagrangian becomes
\br
\lab{mth0}
{\cal L} & = &\frac{1}{2} A_{\mu}A^{\mu} + \alpha
                                                             j^{\mu}_{}\epsilon^{}_{\mu\nu}A^{\nu} - g  j^{\mu}_{}j^{}_{\mu} +
                                                            i\overline{\xi}\gamma^{\mu}_{}\partial^{}_{\mu}\xi 
                                                       - M \overline{\xi}\xi,
\er
where the vector $A_{\mu} \equiv \pa_{\mu} \theta$ has been defined.  One can write the terms containing $A_{\mu}$ as
 \br
\frac{1}{2} A_{\mu}A^{\mu} + \alpha j^{\mu}_{}\epsilon^{}_{\mu\nu}A^{\nu} = \frac{1}{2} (A_{\mu} - \alpha \epsilon_{\mu \nu} j^{\nu})^2 + \frac{\alpha^2}{2} j_{\mu}j^{\mu}\er
Therefore, replacing back this last expression into the Lagrangian (\ref{mth0}) one has 
\br
\lab{mth1}
{\cal L} & = & i\overline{\xi}\gamma^{\mu}_{}\partial^{}_{\mu}\xi 
                                                       - M \overline{\xi}\xi    - \frac{1}{2} g' \,  j^{\mu}_{}j^{}_{\mu} + \frac{1}{2} \pa_{\mu}\sigma \pa^{\mu} \sigma,  
\er
where we have introduced the parameter $g' \equiv -\frac{1}{4} \a^2$ and the scalar field  $\sigma$ through 
\br
\label{sigma10}
\pa_{\mu}\sigma \equiv A_{\mu} - \alpha \epsilon_{\mu \nu} j^{\nu}
\er
So, the final Lagrangian (\ref{mth1}) defines the massive Thirring model for the spinor field $\xi$ with current-current coupling constant  $g'$ plus a free massless scalar field $\sigma$.  
 
Let us discuss the reduction above in the context of the Darboux transformation. In this particular case, i.e.  when $r=\hat{\beta} \rightarrow \beta=0$, one notices that the transformation (\ref{transf1}), supplemented  with $(\pa_{\mu}\theta - \alpha \epsilon_{\mu \nu} j^{\nu}) \rightarrow \pa_{\mu} \sigma$  of (\ref{sigma10}),  becomes truly a Darboux transformation giving rise to the model (\ref{mth1}), which is in the standard canonical representation for the spinor and the free massless scalar field.  In fact, instead of (\ref{atmfj00}) one can get
\begin{eqnarray}
{\cal L} & = &  \frac{1}{2} \partial^{\mu}_{}\theta\partial^{}_{\mu}\theta + 
                                                            \a   j^{\mu}_{}\epsilon^{}_{\mu\nu}\partial^{\nu}_{}\theta - g j^{\mu}_{}j^{}_{\mu} +
                                                            i\overline{\xi}\gamma^{\mu}_{}\partial^{}_{\mu}\xi  
                                                        - M \overline{\xi}\xi \label{atmfj01}
\end{eqnarray}
  
Similarly, as in the procedure above,  one can define  $ \pa_{\mu} \sigma \equiv \pa_{\mu}\theta - \alpha \epsilon_{\mu \nu} j^{\nu} $ and rewrite the Lagrangian (\ref{atmfj01}) as in (\ref{mth1}).

Notice that an alternative symplectic reduction procedure has been performed in \cite{aop}. It has been obtained the massive Thirring sector by setting $r=v=0$ in the initial Lagrangian \rf{lagrangian1}. However, in that process the presence of the free scalar field in the final Lagrangian (\ref{mth1}) did not emerge. As we will see below the free field $\s$ and its trivial solution plays an important role in the understanding of the confining phase of the model. 

\subsection{Sine-Gordon model}

As mentioned above, the Lagrangian (\ref{mthsg}) for $\a = 0$  exhibits the left/right local symmetry of type  (\ref{leri1})-(\ref{leri2}). Therefore,  one must impose a gauge fixing condition to the Lagrangian  (\ref{mthsg}).  So, we will decouple the scalar field $\theta$ from  the spinor degrees of freedom by conveniently gauge fixing this local symmetry   such that
\br 
\label{gfsp1}
\overline{\xi}\xi = -4 \zeta_1^o \zeta_2^o \equiv - \L_o,\,\,\,\,\,\,  \overline{\xi} \g_5\xi =0,\,\,\,\,\,\, j^{\mu} j_{\mu} =0,
\er
where $\zeta_1^o$  and $\zeta_2^o$ are constant Grassmannian parameters. Note that $\L_o$ can be considered as an ordinary commuting real number. The spinor components are defined as $\xi_R = \xi_L =  \zeta_1 + i \zeta_2$, such that  $\zeta_1 \equiv A \zeta_1^o,\,\,\zeta_2 \equiv A^{-1} \zeta_2^o$, with $A$ being an ordinary commuting real function.

So, in (\ref{mthsg}) one sets $\a=0$. In this case one has $ \b \neq 0$. Taking into account these parameters and substituting (\ref{gfsp1}) into the Lagrangian  (\ref{mthsg}) one has
\br
\lab{sg0}
{\cal L} & = &\frac{1}{2} \partial^{\mu}_{}\theta\partial^{}_{\mu}\theta 
                                                       +M \L_{o} \cos{(\beta \theta) }  .                                                            
\er
Notice that  the spinor kinetic terms  contribute a total time derivative  $\sim \frac{d}{dt} \log{A}$ to the Lagrangian, and so, it can be removed.

Then, the F-J symplectic method has been applied to decouple the sine-Gordon and massive Thirring sectors of the model (\ref{atm0}). One can examine the duality correspondence  between these models by inspecting the relationship between the parameters of the model  (\ref{mthsg}). So, from (\ref{abg}) one can write the next relationship between the parameters $\a$ and $\b$    
\br
g' \b^2 &=& -\frac{1}{4}\a^2 \b^2 \\
& \equiv & \d,\,\,\,\,\,\,  \label{dual12}
\er
with
\br
\d \equiv -\frac{4}{\kappa^2} (1- (\frac{r}{\hat{\b}})^2)^2.
\er
Therefore, for $\d = constant$ one can define the strong/weak coupling sectors by examining the relationship (\ref{dual12}). In fact, one has either the strong coupling sector (sine-Gordon model in (\ref{sg0}) with coupling constant $\b$) as $g' \rightarrow 0$ or the weak coupling sector  (Thirring model in (\ref{mth1}) with coupling constant $g'$) as $\b \rightarrow 0$.   

However, it is interesting to analyze the configurations in which the scalar $\theta$ and spinor $\xi$ fields are interacting with intermediate values of the couplings $\a$ and $\b$. So, our results would be relevant to the understanding of the so-called bosonization as duality and smooth bosonization concepts, such that  the bosonization process interpolates smoothly between the bosonic and fermionic sectors of an effective master Lagrangian which describes the coupling of the scalar and the fermion fields \cite{quevedo, smooth}.

\section{Chiral confinement and first order differential equations}

\label{sec:chi}

Next, we will study the properties of the intermediate regions in field space, provided that the coupling parameters satisfy  (\ref{dual12}) with finite and non-vanishing couplings $\{\a, \b\}$. Let us consider the Lagrangian \rf{mthsg} and some of its properties. Due to the conservation law (\ref{chiral}) one can define 
\br
\label{firstorder}
 \epsilon^{\mu\nu}\pa_\nu\Sigma\equiv - \pa^{\mu} \theta + (\alpha+\frac{\beta}{2})\, \bar{\xi}\gamma_5\gamma^{\mu}  \xi  ,  
\er
where we have introduced a new scalar field $\Sigma$. From the last identity one can write 
\br
\label{sigma}
\pa^\mu \Sigma = -\epsilon^{\mu\nu} \pa_{\nu} \theta +  (\alpha+\frac{\beta}{2})\, j^{\mu}, \er
and taking into account the conservation law (\ref{vector}) one has
\br
\label{sig1}
\pa^2 \Sigma = 0.
\er
So, the field $\Sigma$ is a massless free field. Note that (\ref{sigma}) does not define uniquely the field $\S$. In fact, defining  a new field as $\pa^\mu\hat{\S} \equiv  \pa^{\mu}\S + \iota j^{\mu}$\, with  $\iota$ being an arbitrary constant,  one gets another massless scalar free field $\hat{\S}$ satisfying (\ref{sig1}) provided that one assumes the $U(1)$ current  conservation law $\pa_\mu j^\mu =0$ in (\ref{vector}). So, one can write the equations 
\br
\label{sigma1}
\pa^\mu \hat{\S} &=& -\epsilon^{\mu\nu} \pa_{\nu} \theta +  (\alpha+\frac{\beta}{2}- \iota)\, j^{\mu},\\
\pa^2 \hat{\S} &=& 0. \label{sig2}
\er

Taking into account (\ref{sigma1}) the equations of motion (\ref{thetaeq})-(\ref{bpsieq}) become
\br
\label{thetaeq1}
\pa^2 \theta  - \beta (1+\frac{2\a}{\b-2\iota})\, M   \bar{\xi}\xi \sin{(\beta \theta)} - i \beta (1+\frac{2\a}{\b-2\iota})\, M  \bar{\xi}\gamma_5\xi \cos{(\beta \theta)} &=&0\\
\label{psieq1}
i\gamma^{\mu} \pa_{\mu} \xi + 2g \gamma^{\mu} \xi \Big[ (\frac{\alpha}{2g} - \l)\epsilon_{\mu \nu} \pa^{\nu} \theta - \l   \pa_{\mu} \hat{\S}\Big] - M   \xi \cos{(\beta \theta)} +  i M  \gamma_5\xi \sin{(\beta \theta)} &=&0.\\
\label{bpsieq1}
-i \pa_{\mu} \bar{\xi} \gamma^{\mu} + 2 g \Big[ (\frac{\alpha}{2g} - \l)\epsilon_{\mu \nu} \pa^{\nu} \theta - \l   \pa_{\mu} \hat{\S}\Big] \bar{\xi} \gamma^{\mu}  - M   \bar{\xi} \cos{(\beta \theta)} +  i M   \bar{\xi}\gamma_5 \sin{(\beta \theta)} &=&0,
\er
with
\br
\label{l00}
\l & \equiv & \frac{1}{ \alpha+\frac{\beta}{2} - \iota} \,\,.
\er
Remarkably, the set of first order equations (\ref{psieq1})-(\ref{bpsieq1}) for the
Dirac spinors and (\ref{sigma1}) (with $\hat{\S} \equiv \S$) imply the second order differential equation (\ref{thetaeq1}) in the particular case $\iota =0$ (see Appendix \ref{app1}). So, we  expect that in this special case the solutions of the first order system of differential eqs.  (\ref{sigma1})  and (\ref{psieq1})-(\ref{bpsieq1}) will solve the second order differential eq.  \rf{thetaeq1} for the scalar field  $\theta$.

Next, in the special case $\iota =0$, let us consider a trivial solution for the scalar field $\hat{\S}$. So, from \rf{sigma1} one has  
\br
\label{topno1}
 \hat{\S} =0\,\,\,\, \rightarrow\,\,\,\, j^{\mu} &=&\l \, \epsilon^{\mu\nu} \pa_{\nu} \theta.
\er
On can argue that this relationship has been inherited from (\ref{equi11})  upon F-J reduction of the initial model (\ref{atm0}).  Notice the presence of the coupling constants $\alpha$ and $\beta$ through $\l$ in (\ref{l00}) indicating the degree of contribution of them in order to have the equivalence of the Noether and topological currents in the fermion-soliton interacting model. One can argue that the confining sector of the model is determined by the condition $\hat{\S}=0$. This result is in agreement with the quantum field theory result of \cite{npb1}, in which the zero vacuum expectation value of a free scalar field is related to the confining mechanism in the model. 

Notice that setting  $\iota =0$ into the eq. \rf{thetaeq1} one can get
\br
\label{thetaeq1100}
\pa^2 \theta - \frac{2}{\l} M \bar{\xi}\xi \sin{(\beta \theta)} -  \frac{2i}{\l} M  \bar{\xi}\gamma_5\xi \cos{(\beta \theta)} = 0.
\er
Therefore, one can argue that the currents equivalence (\ref{topno1}) gives rise to the change of the coupling strength of the scalar with the spinor bilinears ($ \bar{\xi}\xi $ and $\bar{\xi}\gamma_5\xi$) by a factor  of $\frac{2}{\b \l}$, as compared to the coupling  in (\ref{thetaeq}), at the level of the equations of motion.  

Setting $\hat{\S}=0$ and $\iota=0$ into (\ref{psieq1})-(\ref{bpsieq1}) one gets
\br
\label{psieq1i}
i\gamma^{\mu} \pa_{\mu} \xi + \hat{g}\,  \epsilon_{\mu \nu}\gamma^{\mu} \xi  \pa^{\nu} \theta  - M   \xi \cos{(\beta \theta)} +  i M  \gamma_5\xi \sin{(\beta \theta)} &=&0,\\
\label{bpsieq1i}
-i \pa_{\mu} \bar{\xi} \gamma^{\mu} + \hat{g}\,  \epsilon_{\mu \nu} \pa^{\nu} \theta  \bar{\xi} \gamma^{\mu}  - M   \bar{\xi} \cos{(\beta \theta)} +  i M   \bar{\xi}\gamma_5 \sin{(\beta \theta)} &=&0,
\er
with 
\br
\label{hatg}
\hat{g}  \equiv \frac{1}{8} \big[\frac{(2\a + 5 \b)^2-28\b^2}{2 \a + \b}\big]
\er
So, choosing 
\br
\label{condg}
\hat{g}=0,\er
one can write the system of  equations (\ref{psieq1i})-(\ref{bpsieq1i})  in component form as
\br
\label{xis11}
(\pa_t + \pa_x) \xi_{L}  &=&  -M e^{-i \b \theta}
 \xi_{R}\\
(\pa_t - \pa_x) \xi_{R} &=&  M e^{i \b \theta} \, \xi_{L}, \label{xis12}
\er
plus the complex conjugations of these equations.

Moreover, the first order eqs.  (\ref{topno1}) written in components become
 \br
\label{compo11}
j^{0}&=&\xi^\star_R \xi_R + \xi^\star_L \xi_L = \l\, \pa_{x} \theta,\\
j^{1}&=&-\xi^\star_R \xi_R + \xi^\star_L \xi_L = \l \, \pa_{t} \theta, \label{compo12i}
\er
with
\br
\label{lab1}
\l & =& \frac{1}{ \alpha+\frac{\beta}{2}} = - [\frac{e_1}{1+  e_2 \sqrt{7}}]\, \hat{\b},\\
\a &=&  e_1 [\frac{1-e_2 \sqrt{7}}{2}]\, \hat{\b},\\
\b &=& -e_1 (3+e_2 \sqrt{7})\hat{\b},\,\,\,\,\, e_a = \pm 1,\, a=1,2.  \label{lab11}\er
Note that the first equality of (\ref{lab1}) follows from (\ref{l00}) for $\iota =0$, and the $\a$ and $\b$ formulas (\ref{lab11})  arise from the relationships (\ref{abg}) and the condition (\ref{condg}), i.e. $\hat{g}=0$. Therefore, in the confining regime and with the parameter choice (\ref{condg}) the parameters $\l$, $\a$ and $\b$ in (\ref{lab1})-(\ref{lab11}) appear in terms of the parameter $\hat{\b}$ of the initial Lagrangian (\ref{atm0}). 

From this point onward, we will assume that the parameters satisfy the relationships (\ref{lab1})-(\ref{lab11}). Consequently, the reduced effective model will be determined by the initial parameters $\{M,\hat{\b}\}$ of the Lagrangian (\ref{atm0}).

Remarkably, also in this special case one can consider the first order eqs.  (\ref{xis11})-(\ref{xis12}) 
together with the system (\ref{compo11})-(\ref{compo12i}) as the equations of motion describing the dynamics of the model in the confining phase. Notably,  a direct calculation shows that the system of first order equations  (\ref{xis11})-(\ref{xis12}) and (\ref{compo11})-(\ref{compo12i})  imply the second order eq. for the field $\theta$ (\ref{thetaeq1100}).

Next, we define the system of equations of the reduced ATM model  as
\br
\label{thetaeq1100ra}
\pa^2 \theta - \frac{2}{\l} M \bar{\xi}\xi \sin{(\beta \theta)} -  \frac{2i}{\l} M  \bar{\xi}\gamma_5\xi \cos{(\beta \theta)} &=& 0,\\
\label{psieq1ira}
i\gamma^{\mu} \pa_{\mu} \xi  - M   \xi \cos{(\beta \theta)} +  i M  \gamma_5\xi \sin{(\beta \theta)} &=&0,\\
\label{bpsieq1ira}
-i \pa_{\mu} \bar{\xi} \gamma^{\mu}  - M   \bar{\xi} \cos{(\beta \theta)} +  i M   \bar{\xi}\gamma_5 \sin{(\beta \theta)} &=&0,
\er 
such that the eq. (\ref{thetaeq1100ra}) is the same as (\ref{thetaeq1100}), and the eqs. (\ref{psieq1ira})-(\ref{bpsieq1ira}) come  from (\ref{psieq1i})-(\ref{bpsieq1i}) provided the condition  $\hat{g}=0$ in  (\ref{condg}) is assumed. Note that the system (\ref{thetaeq1100ra})-(\ref{bpsieq1ira}) resembles to the original ATM (\ref{atm0}) eqs. of motion for the scalar and spinor fields, respectively. However, the reduced ATM system (\ref{thetaeq1100ra})-(\ref{bpsieq1ira}) exhibits the effect of the F-J reduction process encoded in the set of coupling parameters $\{\l, \b\}$ defined in (\ref{lab1})-(\ref{lab11}). In fact, the structure of these parameters and their dependence on $\hat{\b}$ arise from the reduction process.   
 
Remarkably, the currents equivalence (\ref{equi11}) as compared to its analog in  (\ref{topno1}) develops a new factor. In fact, the ATM related eq. becomes $\bar{\psi}\g^{\mu}\psi = \sqrt{\frac{\kappa}{2}}\, \epsilon^{\mu \nu}\pa_{\nu} \vp$, whereas the currents equivalence  in the reduced ATM (\ref{topno1}) holds with the $\l$ factor in (\ref{lab1}) as $\l = [\frac{-e_1}{1+e_2 \sqrt{7}}] \sqrt{\frac{\kappa}{2}}$. So, one can argue that the additional factor $[\frac{-e_1}{1+e_2 \sqrt{7}}]$ arises due to the interplay between the parameters $\a$ and $\b$ associated to the spinor-soliton coupling terms present in the reduced ATM model (\ref{mthsg}). So, for the reduced ATM model the eq. (\ref{topno1}) represents the classical equivalence between the Noether and the topological currents. Moreover, it has been shown that, using bosonization techniques,  the initial currents equivalence (\ref{equi11}) holds true at the quantum level, and then reproduces  a bag model like mechanism for the confinement of the spinor fields inside the solitons \cite{npb1}.   

In several nonlinear field theories discovering relevant solutions often involves reducing the order of the original Euler-Lagrange equations. This reduction process simplifies the problem, making it more tractable. For instance, solutions can be found by converting higher-order Euler-Lagrange equations into first-order equations, such as the Bogomolnyi equations,  Backlund transformations and self-duality equations. These first-order equations are easier to solve and can provide significant insights into the underlying physical theories. Some methods in this line have recently been put forward, see e.g. \cite{adam1, ferreira1}. 

In our case the first order system of differential  equations (\ref{xis11})-(\ref{xis12})  and (\ref{compo11})-(\ref{compo12i})  will allow us to find the soliton and bound state solutions of the reduced ATM  model  (\ref{thetaeq1100ra})-(\ref{bpsieq1ira})  in a simpler manner.  

\section{Fermion-kink configurations  and spinor bound states}
\label{sec:bs}

In order to solve the system of equations  (\ref{xis11})-(\ref{xis12})  and   (\ref{compo11})-(\ref{compo12i}) we will use the Hirota tau function approach in which the scalar and the spinor components are parametrized by the tau functions as
\br
\label{tau1f}
e^{\frac{i}{2} \beta \theta} &=& e^{- i \frac{\theta_1}{2}}  \, \,\frac{\tau_1}{\tau_0},\,\,\,\,\, \theta_1 \in \IR, \\
\( \begin{array}{c}
\xi_R \\
\xi_{L} \end{array}\) &=& \sqrt{\frac{m_1}{4 i}} \(\begin{array}{c}
 \tau_R/ \tau_0\\
- \tau_L/\tau_1 \end{array}\),\,\,\,\,\, \( \begin{array}{c}
\xi^\star_R \\
\xi^\star_{L} \end{array}\) = - \sqrt{\frac{m_2}{4 i}} \(\begin{array}{c}
 \widetilde{\tau}_R/ \tau_1\\
\widetilde{\tau}_L/\tau_0 \end{array}\),\label{tau1ff}
\er
with $m_1, m_2$ real parameters.
Substituting the above parametrization into  (\ref{xis11})-(\ref{xis12})  one gets
\br
\label{tau11}
\tau_R (\pa_t - \pa_x)\tau_0 - \tau_0 (\pa_t - \pa_x)\tau_R &=& M \tau_1 \tau_L\\
-\tau_L (\pa_t + \pa_x)\tau_1 + \tau_1 (\pa_t + \pa_x)\tau_L &=& M \tau_0 \tau_R. \label{tau12}
\er
Similarly, substituting into (\ref{compo11})-(\ref{compo12i}) one gets
\br
\label{tau21}
\widetilde{\tau}_R \tau_R - \widetilde{\tau}_L \tau_L  &=& -\frac{ 8\lambda}{\sqrt{m_1 m_2} \beta} (\tau_1 \pa_x \tau_0 - \tau_0 \pa_x \tau_1)\\
\widetilde{\tau}_R \tau_R + \widetilde{\tau}_L \tau_L  &=&  \frac{ 8\lambda}{\sqrt{m_1 m_2} \beta}(\tau_1 \pa_t \tau_0 - \tau_0 \pa_t \tau_1) \label{tau22}
\er

Notice that taking into account 
\br
\tau_1  = (\tau_0)^{\star},
\er 
from (\ref{tau1f}) one can write 
\br
\label{thetatau}
\theta  =  \frac{4}{\b} \arctan{\big\{-i \Big[\frac{e^{-\frac{i}{4} \theta_1} \tau_1 -e^{\frac{i}{4} \theta_1} \tau_0}{e^{-\frac{i}{4} \theta_1} \tau_1 + e^{\frac{i}{4} \theta_1}\tau_0}\Big]\big\}}.
\er
Next, we construct the soliton solutions.

\subsection{Solitons and zero-modes: massive Thirring/sine-Gordon duality}
\label{subsec:1sol}

Let us assume the following expressions for the tau functions for $1-$soliton
\br
\label{tau1}
\tau_1 &=& 1 + \frac{1}{4} e^{-i \theta_o}  a_{+} a_{-} e^{2 \g (x - v t)};\,\,\, \tau_0 = 1 + \frac{1}{4}  e^{i \theta_o}  a_{+} a_{-} e^{2 \g (x - v t)};\\
\label{tau2}
\tau_R &=& \sqrt{i}\, a_{+} z \,e^{\g (x - vt)} ,\,\,\, \widetilde{\tau}_R = \sqrt{i}\, a_{-}\, e^{\g (x - vt)} ,\\
\label{tau3}
\tau_L &=& \sqrt{i}\, a_{+} \, e^{\g (x - vt)} ,\,\,\, \widetilde{\tau}_L = - \sqrt{i}\, \frac{a_{-}}{z} \, e^{\g (x - vt)}.
\er
These expressions solve the system of equations  (\ref{xis11})-(\ref{xis12})  and   (\ref{compo11})-(\ref{compo12i})  provided that the parameters satisfy the relationships
\br
\label{params}
\theta_o &=& - \frac{\pi}{2},\,\,\,\theta_1 = 0,\,\,\,\,\,m_1 m_2  =  \frac{ 16 M^2 \l^2 }{\b^2},\\
\g &=& - \mbox{sign}(z) \frac{M}{\sqrt{1-v^2}},\,\,\,v = \frac{1- z^2}{1+ z^2},\,\,\,\,a_{-} = -  z a^{\star}_{+} \sqrt{\frac{m_1}{m_2}}, \,\,\,\ (a_{+} a_{-}) \in \IR.
\er
Notice that $z$ must be real in order to have a soliton velocity $|v| <1$. So, from (\ref{thetatau}) the kink associated to the field $\theta$ becomes
\br
\label{kak}
\theta_{SG} = \frac{4}{\b} \arctan{[ e^{2 \g (x- v t-x_0)}]},\,\,\,\,\,e^{- 2 \g x_0} \equiv \frac{a_{+} a_{-}}{4}.
\er
This is a 1-soliton type solution of the sine-Gordon model. Even though this soliton for the reduced ATM model  (\ref{thetaeq1100ra})-(\ref{bpsieq1ira}) resembles to the one obtained for the original ATM (\ref{atm0}) scalar field $\vp$ in \cite{npb1}, in the present case the solution (\ref{kak}) encodes an additional source of back-reaction of the spinor $\xi$ on the soliton $\theta$ due to the new coupling constant $\a$, related to $\b$ by  (\ref{lab1}),  which appears in the reduced ATM model (\ref{mthsg}) as the coupling between the $U(1)$ and topological currents. 

Note that the solution (\ref{kak}) exhibits a topological charge 
\br
\label{topok1}
Q_{SG} &=& \frac{\b}{2\pi} (\theta(+\infty)-\theta(-\infty))\\
&=& \pm 1.\label{topok11}
\er 

\begin{figure}
\centering
\label{fig15}
\includegraphics[width=1.5cm,scale=4, angle=0,height=5cm]{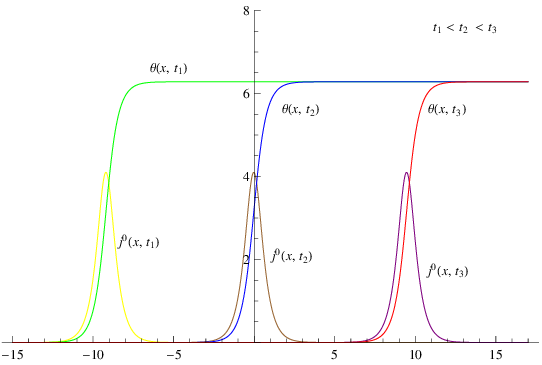} 
\parbox{6in}{\caption{(color online) The kink $\theta(x,t)$ and the confined current component $j^{0}(x,t)$ for successive times $t_1 < t_2< t_3$.}}
\end{figure}  
The $U(1)$ charge density $j^{0}$ associated  to the spinor field  becomes
\br
\label{mth00}
j^0 = \big[\frac{\sqrt{m_1 m_2}}{\cosh{(2\g (x-v t -x_o))}}\big].
\er
In the Fig. 1 we plot the soliton $\theta(x,t)$ and the current component $j^{0}(x,t)$ for three successive times $t_1 < t_2< t_3$. The parameter values are $a_{+}=1,\, z=- 0.8, \, M=1,\, v=0.22,\, m_1=1,\,\b= 1, \l =1.$  
Notice that $j^0$ is significantly confined inside the region of abrupt change of the kink profile associated to the field $\theta$ during the time evolution of the system. So, this plot shows qualitatively the relationship between the zero'th components of the Noether and topological currents equivalence equation  (\ref{topno1}), i.e. it realizes $ \Sigma=0 \rightarrow  j^{0} = \l  \pa_x \theta$. So, this is reminiscent to a bag model like  confinement mechanism of the spinor inside the soliton of this model.

So, in the framework of the Faddeev-Jackiw symplectic method of quantization we have achieved the same picture as the bag model like  confinement mechanism found through the bosonization technique performed in \cite{npb1}.

\subsubsection{Dual sectors and the zero-modes: SG/MT duality}
\label{dual0}

Next, we uncover the dual scalar sector of the zero-modes. For later purpose let us write some identities related to the 1-soliton solution above. The tau functions can be written in terms of the field $\theta$ as follows
\br
\tau_0 &=& \sec{(\frac{\b \theta}{4})} e^{i \b \theta/4} ,\,\,\,\,\,\,\,\,\,\,\,\,\,\,\,\,\,\,\,\,\,\,\,\,\,\tau_1 = \sec{(\frac{\b \theta}{4})} e^{-i \b \theta/4},\\
\tau_R &=& 2 z i \sqrt{\frac{a_{+}}{a_{-}}}\, [ \tan{(\frac{\b \theta}{4})}]^{1/2},\,\,\,\, \widetilde{\tau}_R = 2  i \sqrt{\frac{a_{-}}{a_{+}}} \,[ \tan{(\frac{\b \theta}{4})}]^{1/2},\\
\tau_L &=& 2  i \sqrt{\frac{a_{+}}{a_{-}}} \,[ \tan{(\frac{\b \theta}{4})}]^{1/2},\,\,\,\,\,\,\,\,\widetilde{\tau}_L = -\frac{2}{z}  i \sqrt{\frac{a_{-}}{a_{+}}} \,[ \tan{(\frac{\b \theta}{4})}]^{1/2}.
\er

The current $j^{\mu}$ components become
\br
\label{currcom10}
j^{0} &=& \xi^\star_R \xi_R + \xi^\star_L \xi_L  = \frac{\sqrt{m_1 m_2}}{2} (\frac{1+z^2}{z}) \, \sin{(\frac{\b \theta}{2})},\\
j^{1} &=& -\xi^\star_R \xi_R + \xi^\star_L \xi_L  =  \frac{\sqrt{m_1 m_2}}{2} (\frac{1-z^2}{z}) \, \sin{(\frac{\b \theta}{2})}. \label{currcom20}
\er
The next bilinears will be used below
\br
\label{bil10}
\bar{\xi} \xi  & = & \frac{1}{2}\sqrt{m_1 m_2}  (1 - \cos{(\b \theta)}) , \\
\bar{\xi} \g_5 \xi  & = &  \frac{i}{2} \sqrt{m_1 m_2}   \sin{(\b \theta)},\label{bil20}
\er
and
\br
\label{rl10}
\xi^\star_R \xi_L  &=& - \frac{\sqrt{m_1 m_2}}{4i} (e^{i\b \theta}  - 1 ).
\er
Remarkably, the both dual sectors can be decoupled assuming the above relationships.  So, using (\ref{rl10}) into the eqs. (\ref{xis11})-(\ref{xis12}) one can write
\br
\label{xis11th}
(\pa_t + \pa_x) \xi_{L}  &=& \frac{\b}{\l}\xi^\star_R \xi_L \xi_{R}-i M \xi_R\\
(\pa_t - \pa_x) \xi_{R} &=& \frac{\b}{\l}\xi^\star_L \xi_R \xi_{L} + i M  \xi_{L}, \label{xis12th}
\er
plus the complex conjugations of these equations. This is precisely the system of equations of the massive Thirring model describing the weak coupling sector of the model.

Similarly, taking into account the relationships (\ref{currcom10})-(\ref{currcom20}) into the system (\ref{compo11})-(\ref{compo12i})  one can get
 \br
\label{compo11sg}
 \frac{\sqrt{m_1 m_2}}{2} (\frac{1+z^2}{z}) \, \sin{(\frac{\b \theta}{2})} &=& \l\, \pa_{x} \theta,\\
 \frac{\sqrt{m_1 m_2}}{2} (\frac{1-z^2}{z}) \, \sin{(\frac{\b \theta}{2})} &=& - \l \, \pa_{t} \theta.\,\,\,\,\,\,\,\label{compo12sg}
\er
Notice that this system of first order differential equations  can be written as the following second order equation 
\br
\label{sgeff}
\pa^2 \theta + \frac{4 M^2}{\b} \sin{(\b \theta)} =0.
\er
In fact, this is the sine-Gordon model describing the strong coupling sector of the model. Moreover, using the identities (\ref{bil10})-(\ref{bil20})  into the second order equation for $\theta$ (\ref{thetaeq1100}) one can get the same SG equation (\ref{sgeff}). Remarkably, the 1-kink (\ref{kak}) solves the SG equation (\ref{sgeff}) with the same parameter $\g$ provided by (\ref{params}) with $|v|< 1$.

The non-Hermitian version of the duality mapping above has recently been presented in \cite{jhep24}.

\subsection{1-kink and in-gap fermion bound states}
\label{subsec:1kinkbs}

Let us consider the two-component spinor parametrized as
\br
\label{spbs1}
\xi   = e^{- i \epsilon t}\(\begin{array}c
\xi_R(x)\\
\xi_L(x)\end{array}\).
\er
So, from (\ref{xis11})-(\ref{xis12}) and (\ref{compo11})-(\ref{compo12i}) one can write the coupled system of static equations
\br
\label{sta1}
- i \epsilon \xi_L + \pa_x \xi_L + M e^{-i \b \theta} \xi_R &=&0,\\
\label{sta2}
- i \epsilon \xi_R - \pa_x \xi_R - M e^{i \b \theta} \xi_L &=&0,\\
\xi^\star_R \xi_R + \xi^\star_L\xi_L -  \l \pa_x\theta&=& 0. \label{sta3}
\er
So, the scalar and spinors defined by the relationships (\ref{tau1f})-(\ref{tau1ff}) together with the tau functions (\ref{tau1})-(\ref{tau3}) satisfy (\ref{sta1})-(\ref{sta3}) provided that 
\br
\label{m12}
|z| =1, \,\,\,\,\,\,\, v=0, \,\,\,\,\,\, \theta_1 = - 2 \theta_0,\,\,\,\,\,\,\,
m_1 m_2  =  \frac{ 16 M^2 \l^2 }{\b^2} (\sin{\theta_o})^4,\,\,\,\,\, 
\gamma = M \sin{\theta_o},
\er
with the energy associated to the spinor bound states given by
\br
\label{epsi11}
\epsilon = M \cos{\theta_o} \, \rightarrow \,  \epsilon = \g  \cot{\theta_o},
\er
\br
\label{the11}
\theta_{kink} = -\frac{4}{\b} \arctan{\Big[\tan{(\frac{\theta_o}{2})}\, \tanh{(\g(x-x_0))}\Big]}
\er
and
\br
\label{mth11}
j^0 = \big[\frac{\sqrt{m_1 m_2}}{\cos{\theta_o} + \cosh{(2\g (x-x_o))}}\big].
\er
Some comments are in order here. First, one has that $|\epsilon| \leq M$, then $\epsilon = \pm M$ defines the threshold or half-bound states where the fermion field approaches a constant value at infinity. These type of solutions are finite but they do not decay fast enough at $x= \pm \infty$ to be square integrable \cite{graham}; so,  one can not define a localized charge density $j^0$. 

Second,  considering the normalization condition $\int_{-\infty}^{+\infty} dx j^0=1$ and taking into account (\ref{m12})  one gets 
\br
\label{jo1}
\int_{-\infty}^{+\infty} dx j^0 =1 & \rightarrow &  m_1 m_2  = M^2 \frac{(\sin{\theta_o})^4}{\theta_o^2}\\
\theta_o &=& \frac{1}{8}\b ( 2\a + \b), \label{th0be},\\
\theta_o&=& (\frac{2r_{\pm}+1}{8}) \b^2,\,\,\,\,\,r_{\pm} = -\frac{5}{2}\pm \sqrt{7},\label{thetao2}
\er
where (\ref{th0be}) follows from (\ref{m12}) and the relationship (\ref{lab1}). The eq. (\ref{thetao2}) follows from (\ref{th0be}) and (\ref{lab1})-(\ref{lab11}); so, the $\theta_o$ parameter is proportional to  the square of the coupling constant  $\hat{\b}$, i.e.  $\theta_o \sim \hat{\b}^2$. 
  
Third, for $\theta_o = \pi/2$ one recovers the static version of the zero mode solutions of the subsection \ref{subsec:1sol}. So, the kink soliton (\ref{the11}) is a deformation of the 1-soliton solution of the sine-Gordon model. However, the solution (\ref{the11}) exhibits the topological charge 
\br
\label{fractop}
Q_{kink-top} &=& \frac{\b}{2\pi} (\theta(+\infty)-\theta(-\infty))\\
&=& \frac{2\theta_o}{\pi}, \label{fractop1}
\er 
which is a fractional charge, in contradistinction to the kink/antikink charges in (\ref{topok1})-(\ref{topok11})  which are $\pm 1$ integers, i.e. particular cases of (\ref{fractop1}) for $\theta_o = \pm \pi/2$. Note that the topological charge (\ref{fractop1}) depends on the coupling constant $\b$ according to (\ref{th0be}). This is in contradistinction to the fermion-soliton models studied in the literature, in which the topological charge of the kink has been fixed {\sl a priori}, associated to degenerate vacua of a self-coupling potential of the scalar field sector. In our case the asymptotic behavior of the scalar field and the relevant topological charge is generated dynamically as solutions of the system of first order equations  (\ref{xis11})-(\ref{xis12}) and (\ref{compo11})-(\ref{compo12i}). A model in which quantum effects can stabilize a soliton has been discussed in \cite{farhi}. Our model can be obtained as a particular reduction of the model in \cite{farhi} setting to zero its scalar self-coupling potential. Indeed, our solitons are solutions of \cite{farhi} at the classical level in some region of parameter space.     

Fourth, the form of the relationship (\ref{mth11}) resembles to the one of the usual massive Thirring soliton \cite{han},  provided a convenient parameter identifications are made. In fact, the parameter $\theta_o$ defines the frequency parameter $\omega = \cos{\theta_o}$ of the standing wave soliton solutions of the massive Thirring model. 

Fifth, note that for any value of $\epsilon(\theta_o)$  provided by (\ref{epsi11})  one has  $j^0 =  \l \pa_x \theta$ according to (\ref{epsi11}), i.e. the spinor bound states with energy $\epsilon$ are confined inside the scalar kink.  So, one can argue that the spinor zero-modes are confined inside the kinks which exhibit integer topological charges, whereas the spinor excitation with energy $\epsilon$ becomes confined inside a kink with fractional topological charge.    

Sixth, the soliton-fermion system (\ref{xis11})-(\ref{xis12}) and (\ref{compo11})-(\ref{compo12i}) as a whole can be characterized by two charge densities, the fermionic charge density $j^0(x)$ and the topological charge density defined as the $x-$derivative of the kink $\theta(x)$.

\subsubsection{Dual sectors and the excited states: DSG/dMT duality}
\label{dual1}
We examine the dual sectors of the system of 1-kink and in-gap fermion bound states. Let us write some identities related to the 1-soliton solution for $\theta_o \neq \pm \pi/2$. The tau functions can be written in terms of the field $\theta$ as follows
\br
\tau_0 &=& \sin{\theta_o} \csc{(\frac{\b \theta}{4} + \theta_o )}\, e^{-i \b \theta/4} ,\,\,\,\,\,\,\,\,\,\,\,\,\,\,\,\,\,\,\,\,\,\,\,\,\,\tau_1 = \sin{\theta_o} \csc{(\frac{\b \theta}{4} + \theta_o )}\, e^{i \b \theta/4},\\
\tau_R &=&\pm \sqrt{2}  (1-i) \sqrt{\frac{a_{+}}{a_{-}}}\, \big[ \csc{(\frac{\b \theta}{4} + \theta_o)}\sin{(\frac{\b \theta}{4})}\big]^{1/2},\,\,\,\, \widetilde{\tau}_R = -i \sqrt{\frac{m_1}{m_2}}\, \tau_R^{\star},\\
\tau_L &=& \pm \sqrt{2}  (1-i) \sqrt{\frac{a_{+}}{a_{-}}}\, \big[ \csc{(\frac{\b \theta}{4} + \theta_o)}\sin{(\frac{\b \theta}{4})}\big]^{1/2},\,\,\,\, \widetilde{\tau}_L = i \sqrt{\frac{m_1}{m_2}}\, \tau_L^{\star}.
\er

The current $j^{\mu}$ components become
\br
\label{currcom1}
j^{0} &=& \xi^\star_R \xi_R + \xi^\star_L \xi_L  = \frac{8 M \l}{\b}  \, \sin{(\frac{\b \theta}{4} + \theta_o )} \sin{(\frac{\b \theta}{4})}.
\er
The next bilinears will be used below
\br
\label{bil1}
\bar{\xi} \xi  & = & -\frac{8 M \l}{\b}  \sin{(\frac{\b \theta}{4} + \theta_o )} \sin{(\frac{\b \theta}{4})}  \sin{(\frac{\b \theta}{2})} , \\
\bar{\xi} \g_5 \xi  & = &   -i \frac{8 M \l}{\b}  \sin{(\frac{\b \theta}{4} + \theta_o )} \sin{(\frac{\b \theta}{4})}  \cos{(\frac{\b \theta}{2})},\label{bil2}
\er
and
\br
\label{rl1}
\xi^\star_R \xi_L  &=& -  \frac{4 M \l}{\b} \sin{(\frac{\b \theta}{4} + \theta_o )} \sin{(\frac{\b \theta}{4})}  \, e^{-i\b \theta/2}.
\er
Remarkably, following analogous steps as in the zero-mode case in sec. \ref{dual0}, the above relationships allow us to decouple the scalar and the spinors fields at the level of the equations of motion. So, using (\ref{rl1})  into the eqs. (\ref{sta1})-(\ref{sta2}) one can write
\br
\label{xis11thi}
 - i \epsilon \xi_{L}+\pa_x \xi_{L} - 2 M (\frac{\xi^\star_R \xi_L}{j_0}) \xi_R=0,\\
 - i \epsilon \xi_{R}-\pa_x \xi_{R} - 2 M (\frac{\xi^\star_L \xi_R}{j_0}) \xi_L=0,, \label{xis12thi}
\er
plus the complex conjugations of these equations. This system is a deformation of the massive Thirring model (dMT) describing the weak coupling sector of the reduced ATM model for excited spinor bound states. One can argue that the MT model (\ref{xis11th})-(\ref{xis12th}) has undergone a deformation due to the effect of the kink on the excited spinors in this coupling regime. 

Similarly, taking into account the relationship (\ref{currcom1}) into the equation (\ref{sta3})  one can get
 \br
\label{compo11sgi}
\frac{8 M }{\b} \, \sin{(\frac{\b \theta}{4} + \theta_o )} \sin{(\frac{\b \theta}{4})} &=& \, \pa_{x} \theta.
\er
Notice that this first order differential equation can be written as the next second order equation 
\br
\label{sgeffi}
-\pa_x^2 \theta - \frac{4 M^2}{\b} \sin{(\b \theta + 2\theta_o )} +  \frac{8 M^2}{\b}  \cos{(\theta_o )} \sin{(\frac{\b \theta}{2} +\theta_o )} =0.
\er
Note that for $\theta_o = \frac{\pi}{2}$, corresponding to the zero-mode spinor bound states, this equation reduces to the static version of the SG model (\ref{sgeff}). So, one can argue that (\ref{sgeffi}) is a deformation of the sine-Gordon model (\ref{sgeff}) describing the strong coupling sector of the reduced ATM model for $\theta_o \neq \frac{\pi}{2}$, due to the back-reaction of the excited spinor state with $\epsilon \neq 0$ on the SG soliton. Notice that the relevant Lagrangian associated to this  model possesses the effective potential
\br
\label{pot11}
V_{eff} = \frac{4 M^2}{\b^2} \cos{(\b \theta + 2 \theta_o)} -\frac{16 M^2 \cos{\theta_o}}{\b^2} \cos{(\frac{\b \theta}{2} + \theta_o )}. 
\er
This potential defines the non-integrable double SG model (DSG). Notice that the topological kinks may interpolate two neighboring points of the vacua $\{\frac{-4\theta_o}{\b},0, \frac{4(\pi-\theta_o)}{\b}\}$ of the potential $V_{eff}$. This potential is plotted in the Fig. 2.

A 1-kink  solution of the  DSG equation (\ref{sgeffi}) becomes
\br
\label{defSG1}
\theta_{DSG}=\d_1  \frac{4}{\b} \arccos{\Big\{ \d_2 \,\frac{\mbox{sech}(2 M x \sin{\theta_o})-\cos{\theta_o}[1+\tanh{(2 M x \sin{\theta_o})}] }{\sqrt{2} \sqrt{[1-\cos{\theta_o} \mbox{sech}(2 M x \sin{\theta_o})][1+\tanh{(2 M x \sin{\theta_o})}] }}\Big\}},\,\,\,\,\, \d_{1,2} = \pm 1. \er
The kinks $\theta_{DSG}$ in the Figs 3 and 4 (red lines) show this solution for certain values of $\theta_o$ and they  interpolate the points $\{0, \frac{4(\pi-\theta_o)}{\b}\}$ of the vacua mentioned above. Their topological charges can be defined as
\br
\label{deftop}
Q_{DSG-top} &=& \frac{\b}{2\pi} (\theta(+\infty)-\theta(-\infty))\\
&=&\pm 2 (1-\frac{\theta_o}{\pi}), \label{fractop1d}
\er 

\begin{figure}
\centering
\label{fig1k}
\includegraphics[width=1.5cm,scale=4, angle=0,height=5cm]{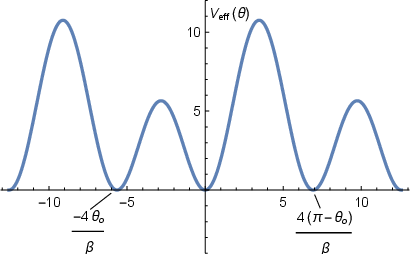} 
\parbox{6in}{\caption{(color online) Plot of the potential $V_{eff}$. Notice the vacua $\{\frac{-4\theta_o}{\b},0, \frac{4(\pi-\theta_o)}{\b}\}$. The kinks $\theta_{SG}$ in the Figs 3 and 4 (red lines) interpolate the points $\{0, \frac{4(\pi-\theta_o)}{\b}\}$ of this vacua.}}
\end{figure} 

\begin{figure}
\centering
\label{fig11k}
\includegraphics[width=1.5cm,scale=4, angle=0,height=4cm]{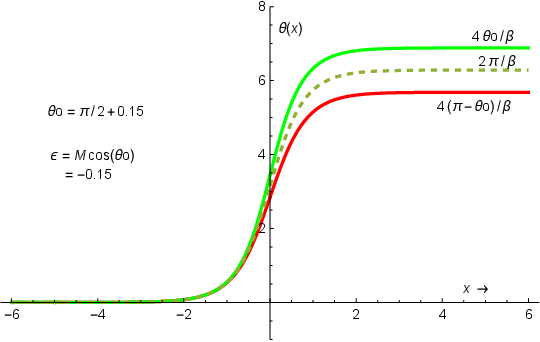} 
\includegraphics[width=1.5cm,scale=4, angle=0,height=4cm]{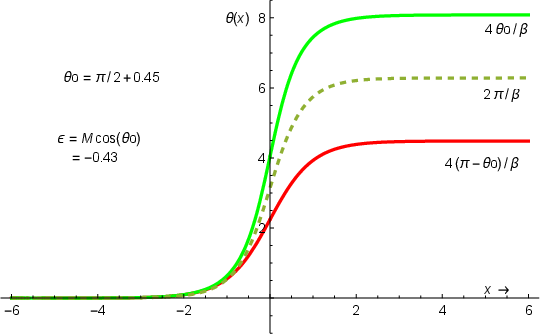} 
\parbox{6in}{\caption{(color online) The SG soliton $\theta_{SG}$ (dashed) (\ref{kak}), the kink $(\theta_{kink}+ \frac{2 \theta_o}{\b})$ (green) (\ref{the11}) and the strong coupling deformed SG $\theta_{DSG}$   (red)  (\ref{defSG1}). For $\b=1$, $\theta_{o}=(\frac{\pi}{2}-0.16),\,M=1,\, \d_2=\d_1=1.$ For these values notice the positive value of the spinor excitation energy $\epsilon=+0.16.$}}
\end{figure}

\begin{figure}
\centering
\label{fig2k}
\includegraphics[width=1.5cm,scale=4, angle=0,height=4cm]{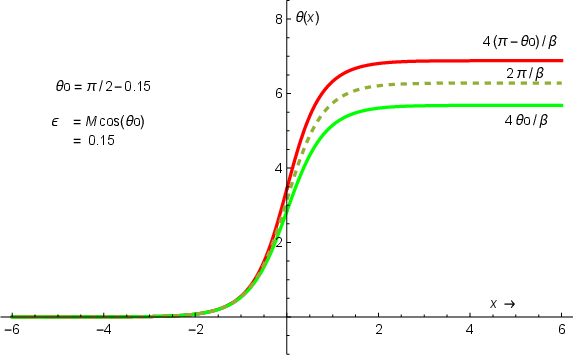} 
\includegraphics[width=1.5cm,scale=4, angle=0,height=4cm]{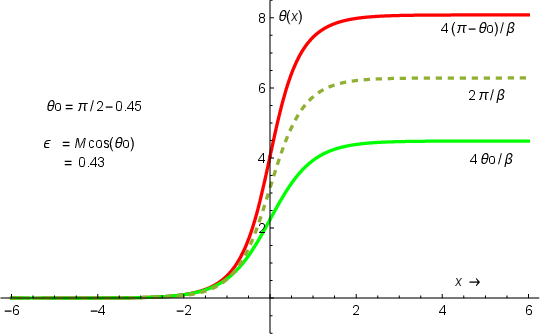} 
\parbox{6in}{\caption{(color online) The SG soliton $\theta_{SG}$ (dashed) (\ref{kak}), the kink $(\theta_{kink}+ \frac{2 \theta_o}{\b})$ (green) (\ref{the11}) and the strong coupling deformed SG $\theta_{DSG}$   (red)  (\ref{defSG1}). For $\b=1$, $\theta_{o}=(\frac{\pi}{2}+0.16),\,M=1,\, \d_2=\d_1=1.$ For these values notice the negative value of the spinor excitation energy $\epsilon=-0.16.$}}
\end{figure} 
Note that the DSG soliton  (\ref{defSG1}) defined for $\theta_o \neq \{0 ,  \pi\}$ exhibits fractional topological charges in  (\ref{fractop1d}). The case $\theta_o =  0$ with $Q_{DSG-top} =  2$ corresponds to the threshold spinor bound state ($\epsilon =  M$) and, as discussed above, they do not correspond to localized spinor charge densities.  
 
The Figs. 3 and 4 show  the relevant kinks. Note that their profiles and asymptotic values are related to the  $\theta_o$ parameter, as well as the relevant excited spinor energies. In fact, one notices that the topological charge of the relevant kink will depend on the value of $\theta_o$, and then on the bound state energy due to the relationship $\epsilon(\theta_o) = M \cos{\theta_o}$. The SG kink $\theta_{SG}$ (dashed) in the both Figs.  possesses a topological charge equal to unity, since this kink corresponds to the spinor zero-mode $\epsilon=0$.  

Comparing the left and right panels of the Fig. 3 one notices that the asymptotic values $\theta_{kink}(+\infty)$ [for fixed $\theta_{kink}(-\infty)=0$]  for the kink $\theta_{kink}$ (green) increases as the value of $\epsilon$ decreases from $-0.15$(left panel) to $-0.43$(right panel). So, its relevant topological charge increases. However, the asymptotic value of the decoupled DSG kink $\theta_{DSG}(+\infty)$ [$\theta_{kink}(-\infty)=0$] (red) decreases as $\epsilon$ decreases; so,  its topological charge  decreases as $\epsilon$ decreases. 

Likewise, comparing the left and right panels of Fig. 4 one notices that the asymptotic values $\theta_{kink}(+\infty)$ [for fixed $\theta_{kink}(-\infty)=0$]  for the kink $\theta_{kink}$ (green) decreases as the value of $\epsilon$ increases from $0.15$(left panel) to $0.43$ (right panel). So, its relevant topological charge decreases. On the other hand, the asymptotic value of the decoupled DSG kink $\theta_{DSG}(+\infty)$ [$\theta_{kink}(-\infty)=0$] (red) increases as $\epsilon$ increases; so,  its topological charge  increases as $\epsilon$ increases. 

Remarkably, from the behavior in the both Figs. one can conclude that the topological charge of the kink $\theta_{kink}$ (green) increases as $\epsilon$ decreases; so, the system exhibits a  topological charge pumping mechanism as the effect of the back-reaction of the spinor bound state on the kink. One can argue that this mechanism, driven by the fermionic back-reaction, exhibits the dynamic interplay between fermionic excitations and topological features, offering new insights into the non-trivial topology of integrable models and their deformations. In fact, the non-integrale DSG model potential (\ref{pot11}) represents a deformation of the integrable SG model. 

\section{Energy of kink-fermion configuration plus spinor bound states}
\label{sec:energy}

In this section we compute the energy of the soliton-fermion configurations plus the excited fermion bound state energy $\epsilon$, associated to the reduced ATM model 
(\ref{thetaeq1100ra})-(\ref{bpsieq1ira}). We perform this computation firstly by writing the energy density associated to the Lagrangian (\ref{mthsg}) for static configurations, and then specializing the result for the on-shell first order system of equation (\ref{sta1})-(\ref{sta3}), with the set of parameters (\ref{lab1})-(\ref{lab11}), (\ref{epsi11}) and (\ref{th0be}). So, from (\ref{mthsg}) one can define
\br
{\cal H} = \dot{\theta} \Pi_{\theta} + \dot{\xi}_R \Pi_{R} +  \dot{\xi}_L \Pi_{L}  - {\cal L},
\er
with
\br
\Pi_{\theta} \equiv  \dot{\theta} + \a j^{1} ,\,\,\, \Pi_{R} \equiv -i \xi^\star_R,\, \,\,\,  \Pi_{L} \equiv -i \xi^\star_L.
\er
Therefore, one has
\br
\nonumber
H  &=& \int_{-\infty}^{\infty}\, dx\, {\cal H}\\
\nonumber
      &=&  \int_{-\infty}^{\infty}\,  \, dx \Big\{  \frac{1}{2}[\theta^{'} - (\a+ \frac{\b}{2}) j^0]^2 + \frac{\b}{2} j^{0}[\theta^{'} - (\a+ \frac{\b}{2}) j^0] + \frac{1}{2}\Pi_{\theta}[\Pi_{\theta} - (\a -\frac{1}{\l}) j^{1}] \\
\nonumber
&& - \Pi_{R}  [\pa_x \xi_R + M  e^{-i \b \theta} \xi_L] + \Pi_{L}  [\pa_x \xi_L + M e^{i \b \theta} \xi_R] \\
\label{HH1}
&& -\frac{1}{32} [(2 \a + \b)^2+8\b (\a -  \b)] (j^{0})^2 - (g-\frac{\a^2}{2}) (j^{1})^2 - \frac{1}{2} (\frac{1+ \l \a}{\l}) \Pi_{\theta} j^{1}\Big\} 
\er
The energy of static configurations (set $\Pi_{\theta} = \a j^{1}$)  can be written as
\br
\nonumber
E  &=&  \int \, dx \Big\{  \frac{1}{2}[\theta^{'} - (\a+ \frac{\b}{2}) j^0]^2 +\frac{\b}{2} j^{0}[\theta^{'} - (\a+ \frac{\b}{2}) j^0]\\
\nonumber
&&   - \Pi_{R}  [\pa_x \xi_R + M  e^{i \b \theta} \xi_L] + \Pi_{L}  [\pa_x \xi_L + M e^{-i \b \theta} \xi_R] -\\
&&   \frac{1}{32} [(2 \a + \b)^2+8\b (\a -  \b)] (j^{0})^2+\frac{1}{35}[52 \a^2 +28 \a \b -3 \b^2] (j^{1})^2 \Big\}. \label{en0}
\er
In order to compute $E$ we  assume the static field configurations satisfy the first-order equations (\ref{sta1})-(\ref{sta3}). For static soliton-fermion solutions one has $j^{1} =0$, since $j^{1} = -\l \pa_t \theta$  from (\ref{compo11}). As an explicit realization of this one can notice that the requirement $z=\pm 1$ implies $v=0$  in (\ref{params}) and also $j^{1}=0$ in (\ref{currcom20}). So, the energy (\ref{en0}) of the static configurations becomes  
\br
\label{en11}
E =  -\frac{1}{32} [(2 \a + \b)^2+8\b (\a -  \b)] \int_{-\infty}^{+\infty} \, (j^{0})^2 dx + \epsilon\, \int_{-\infty}^{+\infty} j^{0} dx.  
\er  
Notice that the first two terms of (\ref{en0}) vanish identically upon using the first order eq. (\ref{sta3}) with $\l = \frac{1}{\a+\b/2}$; i.e.  $j^{0} = \l\, \theta'$. The last term in  (\ref{en11}) arises  upon using the first order equations (\ref{sta1})-(\ref{sta2}) into the terms of the second line of (\ref{en0}). Taking into account the static form of (\ref{compo11}), or equivalently (\ref{sta3}), the expression (\ref{en11}) can be written as
 \br
\label{en1}
E &=&  -\frac{1}{32} [(2 \a + \b)^2+8\b (\a -  \b)] \int_{-\infty}^{\infty} dx \, j^{0}  \l  \pa_x \theta \,\, + \epsilon\, \int_{-\infty}^{+\infty} j^{0} dx,\\
\label{en2}
 &=&  -\frac{1}{32} [(2 \a + \b)^2+8\b (\a -  \b)] \l \int_{\theta_1}^{\theta_2} d\theta \, j^{0}\,\, + \epsilon\, \int_{-\infty}^{+\infty} j^{0} dx,\\
 &=&   -\frac{1}{32} [(2 \a + \b)^2+8\b (\a -  \b)] \l  \big[ {\cal J}(\theta_2)-{\cal J}(\theta_1)\big] \,\,  + \epsilon, \label{en3}
\er
where it has been used the identity $(j^{0})^2 = j^{0}  \l  \pa_x \theta  $ and a pre-potential $\cal{J}(\theta)$ has been defined as 
\br
\label{pretpot0}
 j^{0} \equiv \frac{d}{d\theta} \cal{J}(\theta),
\er
and the normalization condition $\int_{-\infty}^{+\infty} j^{0} dx =1$ has been used in the last integral. Then the first term in  (\ref{en3}) represents the energy of the kink-fermion configuration and the last term the excited energy $\epsilon$ of the spinor bound state.

The first integration in (\ref{en2}) is between the two neighboring points $\theta_1$ and $\theta_2$ between which the soliton $\theta$ interpolates. We assume the current component $j^{0}$ and the pre-potential $\cal{J}(\theta)$ to be related to a spinor bound state and a topological soliton which obeys non-trivial boundary conditions. For example, the field $\theta$ supports topological solitons in the form of a kink (\ref{the11}) coupled to a spinor bound state (\ref{spbs1}) with energy $\epsilon = M \cos{\theta_o}$ in (\ref{epsi11}) and charge density (\ref{mth11}).  

So, in order to compute the energy $E$ in (\ref{en3}) of the whole soliton-spinor configuration plus $\epsilon$ it suffices to know the two asymptotic values of the soliton $\theta$ and the parameter value $\theta_o$. Below we compute the energy of the soliton-spinor configurations provided by the SG soliton $\theta_{SG}$ (\ref{kak}) and the related zero-mode fermion with charge density (\ref{mth00}), as well as  the kink  $\theta_{kink}$ (\ref{the11}) coupled to its relevant excited spinor bound sate with charge density (\ref{mth11}), respectively. Moreover, in order the compare with $E$ we compute the energy $E_{DSGkink}$ of a decoupled scalar DSG $\theta_{DSG}$ kink (\ref{defSG1}).  

{\bf 1. Spinor zero-mode coupled to SG soliton $\theta_{SG}$ (\ref{kak})}. From (\ref{pretpot0}) and taking into account (\ref{currcom10}) with $z=1$, one has  
\br
\label{prepot0}
{\cal J}_{SG} =  -\frac{8 M \l}{\beta^2} \cos{(\frac{\b \theta}{2})}.
\er
Therefore, setting $\theta_2= \frac{2\pi}{\b}, \theta_1=0$ into (\ref{en3}) one has 
\br
\label{SGener}
E_1 &=&  -\,\frac{2 M (2 \a -\b)(2\a + 7\b)}{\b^2 (2 \a +\b)^2}. \\
&=& \frac{(2 r_{\pm} -1)(2 r_{\pm} + 7)}{(2 r_{\pm} +1)^2}  \frac{2 M}{\b^2},\,\,\,\,\, r_{\pm} \equiv -5/2\pm \sqrt{7},\,\,\ \a = \b  r_{\pm}.\label{SGener1}
\er 
The last term $\epsilon$ in (\ref{en3}) vanishes since in this case one has the zero mode for $\theta_o= \frac{\pi}{2}$ in (\ref{epsi11}). Notice that $E_1$ represents the energy of the soliton-fermion system and satisfies $|E_1| < E_{SG}$, where $E_{SG} =  \frac{16M}{\b^2}$ is the energy of the static soliton of the SG model which can be computed for the kink (\ref{kak}) as $E_{SG} = \int_{min} d\theta  \sqrt{2 V}$. Moreover, $E_1$ in (\ref{SGener1}) has been written in terms of the parameters $\{M, \b\}$, since $\a = \b  r_{\pm}$ from the parameter relationship (\ref{lab1}).   

{\bf 2. Energy of kink-fermion configuration plus spinor bound state}. One considers the kink  (\ref{the11}) coupled to the spinor bound state with energy $\epsilon$. Using the identity  (\ref{currcom1}) and the definition (\ref{pretpot0}) one can write
\br
\label{prepot1}
{\cal J}_{kink} =  \frac{4 M \l}{\beta^2} \,  [\b \theta \cos{\theta_o}- 2 \sin{(\frac{\b \theta}{2} + \theta_o)}].
\er
Then, inserting this last relationship into (\ref{en3}) with $\theta_2=\frac{4\theta_o}{\b},\,\theta_1=0$ one has the energy of the kink-fermion plus the spinor bound state configurations as   
\br
\label{E2}
E &=& E_{kf} + \epsilon\\
E_{kf} &\equiv& -M \,\frac{(2 \a -\b)(2\a + 7\b)}{\b^2 (2 \a +\b)^2} \,\big[2 \, \theta_o \cos{\theta_o} + \sin{\theta_o} -  \sin{(3 \theta_o)}\big],\,\,\,\,\,\,\,\,\,\epsilon \equiv  M \cos{\theta_o}. \label{E2i}
\er
Notice that in the zero-mode case inserting $\theta_o = \pi/2$ into (\ref{E2}) one recovers the energy $E_1$ in (\ref{SGener}).  As in the literature \cite{weigel1, weigel2, lee} we call $E$ (\ref{E2}) the {\sl quasi-classical} energy. 

{\bf 3. Decoupled DSG kink (\ref{defSG1}) energy} 

Using $E=\int_{\theta_1}^{\theta_2} d\theta \sqrt{2V}$  with $\theta_2=\frac{4\pi}{\b}(1- \frac{\theta_o}{\pi})$\,and $\theta_1=0$ corresponding to the asymptotic values of the kink (\ref{defSG1}) one has the energy  
\br
\label{E31}
E_{DSGkink} = \frac{16 M}{\b^2}\,  \big[(\pi-\theta_o) \cos{\theta_o}+ \sin{\theta_o} \big].
\er 
Notice that setting  $\theta_o = \pi/2$ into (\ref{E31}) one recovers the energy of the static soliton of the SG model, $E_{SG} =  \frac{16M}{\b^2}$. Moreover, setting  $\theta_o = 0$ in (\ref{E31}) one recovers the energy of the static soliton of the DSG model, $E_{DSG} =  \frac{16M \pi}{\b^2}$, corresponding to the threshold bound state energy $\epsilon = M$.

In the Figs. 5 and 6 we present the various energies plotted as functions of the coupling constant $\b$. The top left panels show $E\, vs\, \b$ (red) and $E_{DSGkink} \, vs\, \b$ (green). In the top left panels the dashed lines show the threshold values $\epsilon= \pm 1$. The top right panels shows $E_{kf}\, vs\, \b$ (brown). The bottom left ones show $\epsilon\, vs\, \b $ (magenta) and the bottom right ones $(E_{kf} - \epsilon)\, vs \, \b$ (blue). The Figs. are plotted for $M=1, \a = (r_{\pm})\b\,\, (r_{\pm} =-5/2\pm \sqrt{7})$ (Figs. 5 for $r_+$ and Fig. 6 for $r_-$). From the bottom right Figs. one can argue that for $\b \rightarrow$ large, one has $E_{kf} \approx \epsilon$.

\begin{figure}
\centering
\label{fig3k}
\includegraphics[width=1.5cm,scale=4, angle=0,height=4cm]{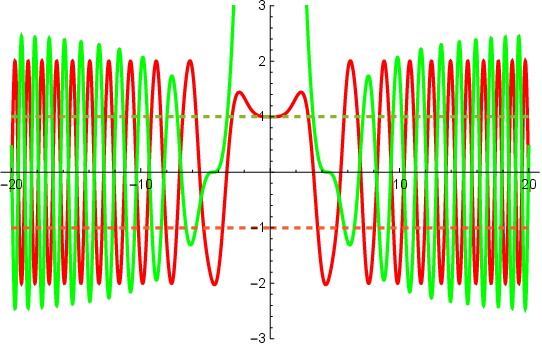} 
\includegraphics[width=1.5cm,scale=4, angle=0,height=4cm]{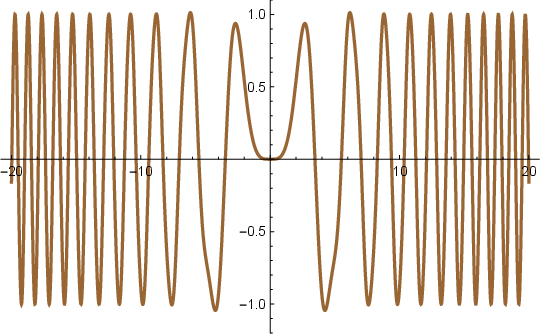}
\includegraphics[width=1.5cm,scale=4, angle=0,height=4cm]{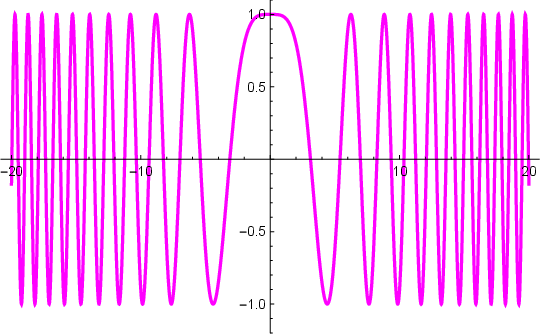}
\includegraphics[width=1.5cm,scale=4, angle=0,height=4cm]{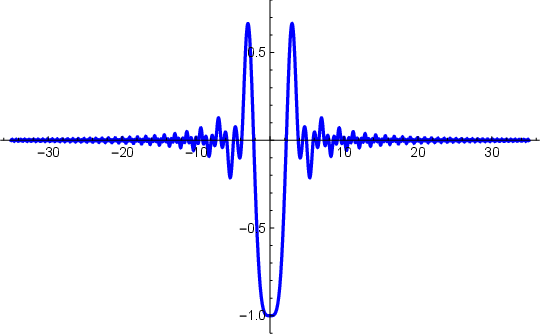}
\parbox{6in}{\caption{(color online) The top left panel show $E\, vs\, \b$ (red) and $E_{DSGkink} \, vs\, \b$ (green). In the top left the dashed lines show the threshold values $\epsilon= \pm 1$. The top right shows $E_{kf}\, vs\, \b$ (brown). The bottom left shows $\epsilon\, vs\, \b $ (magenta) and the bottom right $(E_{kf} - \epsilon)\, vs \, \b$ (blue). The bottom right shows that for $\b \rightarrow large$ one has $E_{kf} \approx \epsilon$. The Figs. are plotted for $M=1, \a = (r_{+})\b\,\, (r_{+} =-5/2+\sqrt{7} \approx 0.15) $.}}
\end{figure}

\begin{figure}
\centering
\label{fig4k}
\includegraphics[width=1.5cm,scale=4, angle=0,height=4cm]{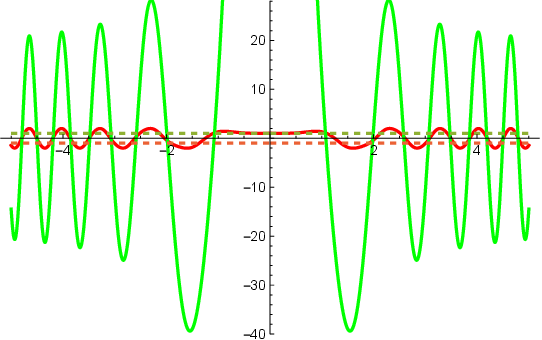} 
\includegraphics[width=1.5cm,scale=4, angle=0,height=4cm]{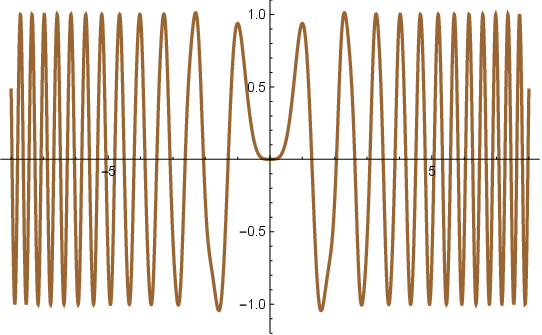}
\includegraphics[width=1.5cm,scale=4, angle=0,height=4cm]{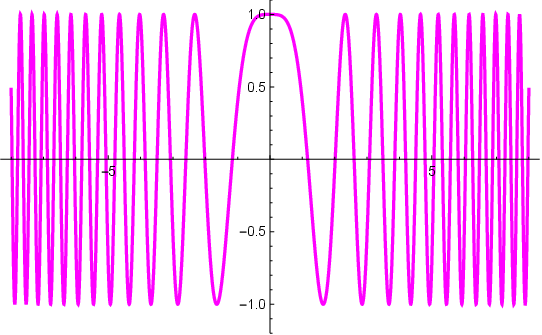}
\includegraphics[width=1.5cm,scale=4, angle=0,height=4cm]{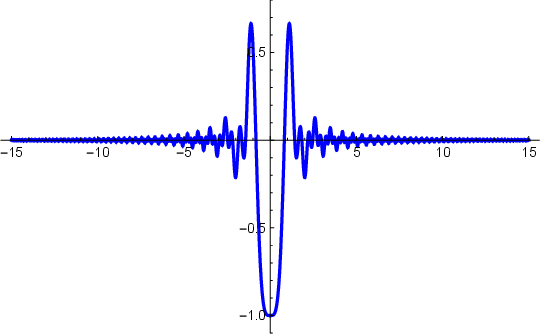}
\parbox{6in}{\caption{(color online) The top left panel show $E\, vs\, \b$ (red) and $E_{DSGkink} \, vs\, \b$ (green).  In the top left the dashed lines show the threshold values $\epsilon= \pm 1$. The top right shows $E_{kf}\, vs\, \b$ (brown). The bottom left shows $\epsilon\, vs\, \b $ (magenta) and the bottom right $(E_{kf} - \epsilon)\, vs \, \b$ (blue). The bottom right shows that for $\b \rightarrow large$ one has $E_{kf} \approx \epsilon$. The Figs. are plotted for $M=1, \a = (r_{-})\b\,\, (r_{-} =-5/2-\sqrt{7} \approx - 5.15) $.}}
\end{figure} 

In the Figs. 5 and 6 one has $M=1$. In the top right panels one has $|E_{kf}|  \leq 1$, in the bottom left panels $|\epsilon|  \leq 1$, whereas in the top left panels $|E| \leq 2$; so they reproduce the relationship $|E| = |E_{kf} + \epsilon| \leq 2$. Notably, one has a kink-fermion configuration energy $E_{kf}$,  as well as  a normalized number one bound state energy $\epsilon$, whose energy values are below that of a single free fermion. 
 
In the top left panels the decoupled  DSG kink energy satisfy $|E_{DSGkink}| > 1$ for some regions of the $\b$ coupling parameter (green). This last relationship implies the existence of kinks with energies above and below the threshold energies $\epsilon= \pm 1$, i.e. stable kink states lying in the continuum of scattering states (KIC states). This is in contradistinction to the bound states in the continuum (BIC states) present in the ATM model plus a scalar self-coupling potential recently studied in \cite{jhep22}.            

The system of first-order equations (\ref{sta1})-(\ref{sta3}) serves a comparable role to the Bogomolnyi-Prasad-Sommerfeld (BPS) equations, as they not only yield the second-order Euler-Lagrange equation for the scalar field but also determine the total energy (\ref{en3}) based on the parameters of the topological charges. In fact, the first-order differential equations (\ref{sta1})-(\ref{sta3}) are essential in relation to our energy functional (\ref{en0}) and the static energy (\ref{en11}). The BPS bounds are a powerful tool for finding topological soliton solutions because they impose constraints on soliton energies based on a topological charge. Solitons that reach this bound must satisfy specific first-order differential equations, known as BPS equations.  

To derive the first-order equations, we reduced the order of the chiral current conservation equation (\ref{chiral}) by introducing a massless free field, $\Sigma$, as shown in equations (\ref{firstorder})-(\ref{sig1}). Within this framework, the trivial solution $\Sigma =0$ leads to the first-order equation (\ref{topno1}). So, it parallels the approach proposed in \cite{devega}, where the authors derived first-order equations for vortices in 1+2 dimensions by considering the conservation of the energy-momentum tensor. Our method differs from the Bogomolnyi trick, which obtains first-order equations by completing the square in the energy functional. However, our approach is similar to the BPS method in that it expresses the total energy in terms of the asymptotic values of the scalar field which are related to the topological charges. 

\section{Dirac sea modification due to the soliton}
\label{sec:diracvacuum}

The modification of the fermionic energy spectrum induced by the presence of the soliton occurs because the soliton alters the fermionic field modes, generating bound states and scattering states distinct from those in a free system. The energy contribution due to the interaction between the kink and the Dirac sea is essential for ensuring the consistency of the semi-classical expansion in the fermionic sector. Below we will compute the scattering states of the fermion-soliton ATM model. 

Let us consider the two-component spinor parameterized as
\br
\label{spbs11}
\xi   = e^{- i E_1 t}\(\begin{array}c
\zeta_R(x)\\
\zeta_L(x)\end{array}\),
\er
where the spinor components $\zeta_R(x)$ and $\zeta_L(x)$ define the scattering solutions in the presence of the soliton $\theta$, and $E_1$ is the energy of these states. 
 
So, from (\ref{xis11})-(\ref{xis12}) and (\ref{compo11})-(\ref{compo12i}) one can write the coupled system of static equations
\br
\label{sta11}
- i E_1 \zeta_L + \pa_x \zeta_L + M e^{-i \b \theta} \zeta_R &=&0,\\
\label{sta21}
- i E_1 \zeta_R - \pa_x \zeta_R - M e^{i \b \theta} \zeta_L &=&0,\\
\Big\{\zeta^\star_R \zeta_R + \zeta^\star_L\zeta_L - \Big[\zeta^{\star\,(free)}_R \zeta^{\,(free)}_R + \zeta^{\star\,(free)}_L\zeta^{(free)}_L\Big](x = - \infty)\Big\} -  \l \,\pa_x\theta&=& 0, \label{sta31}
\er  
where the symbol $^{\star }$ stands for complex conjugation as usual. 

Note that in (\ref{sta31}) we have subtracted the contribution of the charge density due to the free state $\zeta^{\,(free)}_{R, L}$ evaluated at $x = - \infty$ to align it with the scalar field derivative. Specifically, equation (\ref{sta31}) is consistent when applied to the asymptotic regions of the soliton, where the derivative of the kink-like scalar field vanishes, and the spinor fields reduce to free Dirac plane waves, i.e. $ \zeta_{R, L} \rightarrow \zeta^{\,(free)}_{R, L}$. This will imply a further relationship between the parameters of the solution evaluated at  $x = + \infty$, as we will see below. Furthermore, we will explore the scattering solutions that satisfy (\ref{sta31}) across the entire real line.

For plane waves, the charge density term inside square bracket denoted by the superscript `free' in (\ref{sta31}) reduces to a constant. Consequently, even in the case of scattering states interacting with a soliton field, it can be argued that the first-order system of equations (\ref{sta11})-(\ref{sta31}) also leads to the second-order differential equation for the scalar field (i.e. the static version of the eq. (\ref{thetaeq1100ra})), as previously discussed for the bound states. 

Next, we compute the spinor states scattering from the soliton of type (\ref{tau1f}). So, let us consider the tau functions 
\br
\label{stau0}
\tau_0 &=& 1+ \frac{1}{4} e^{-i \theta_o} c_0  e^{2 k x},\\
\label{stau1}
\tau_1 &=& 1+ \frac{1}{4} e^{i \theta_o} c_0  e^{2 k x}, \\
\label{staur}
\tau_R &=& d^{+}  e^{2 k x}, \,\,\,\,\, \widetilde{\tau}_R = d^{+\star}\,  e^{2 k x},\\
\label{staul}
\tau_L &=& d^{-}  e^{2 k x}, \,\,\,\,\, \widetilde{\tau}_L = d^{-\star} \, e^{2 k x}.
\er
Taking into account (\ref{tau1f}) and  (\ref{thetatau}) one can write the scalar soliton as
\br
\label{solsc}
\theta = \frac{4}{\b} \arctan{ \Big\{ -\tan{(\frac{\theta_1}{4})} \Big[ \frac{1+ c_0 \frac{\sin{(\frac{\theta_1}{4}+ \theta_0)}}{\sin{(\frac{\theta_1}{4})}} e^{2 k x}}{1+ c_0 \frac{\cos{(\frac{\theta_1}{4}+ \theta_0)}}{\cos{(\frac{\theta_1}{4})}} e^{2 k x} }\Big] \Big\}  }.
\er
Note that this solution reproduces the soliton (\ref{the11}) provided that $\theta_1= - 2 \theta_0$. The relevant asymptotic values  of the soliton (\ref{solsc}) define the quantity 
\br
\label{toposc1}
\D \theta  &=&  \theta(x\rightarrow +\infty)-\theta(x\rightarrow -\infty ) \\
&=& \left\{\begin{array}{cr} - \frac{4\theta_o}{\b} &  k>0\\
\frac{4 \theta_o}{\b} &  k<0
\end{array}\right.
\label{toposc2}
\er 

Let us consider the scattering states as  
\br
\label{ztau1}
\zeta_R &=& e_R e^{i k_1 x} + e^{i k_1 x} \frac{\tau_R}{\tau_0},\,\,\,\,\, \zeta^{\star}_R = e^\star_R e^{-i k_1 x} + e^{-i k_1 x} \frac{\widetilde{\tau}_R}{\tau_1},\\
\label{ztau2}
\zeta_L &=& e_L e^{i k_1 x} + e^{i k_1 x} \frac{\tau_L}{\tau_1},\,\,\,\,\, \zeta^{\star}_L = e^\star_L e^{-i k_1 x} + e^{-i k_1 x} \frac{\widetilde{\tau}_L}{\tau_0}.
\er
It is important to note that the solution (\ref{ztau1})–(\ref{ztau2}) describes a wave state propagating over the free spinor continuous background, where the wave numbers $k_1$ and $k$ correspond to the contributions from the free field and the soliton-induced component, respectively. Below, we will establish a relationship between them. Similar scattering states  have been considered in the search for self-consistent solutions of complex-valued fermionic condensates in the (1+1)-dimensional Bogoliubov-de Gennes equation (BdG) \cite{takahashi1}. In fact, the spinor sector of the ATM model (\ref{sta11})-(\ref{sta21}) is similar to the BdG equation with gap function being a pure phase.  

The wave functions (\ref{ztau1})-(\ref{ztau2}) exhibit the
asymptotic forms
\br
\left(\begin{array}{c}
\zeta_R \\
\zeta_L
\end{array}\right) &\xrightarrow[x \rightarrow -\infty]{\,}& \left(\begin{array}{c}
e'_R\, e^{-i k_1 x} + e_R\, e^{i k_1 x} \\
e'_L\, e^{-i k_1 x} + e_L\, e^{i k_1 x} 
\end{array}\right),\,\,\,\,\,\,\, e'_R = e'_L =0, \label{left1}\\
\left(\begin{array}{c}
\zeta_R \\
\zeta_L
\end{array}\right) &\xrightarrow[x \rightarrow +\infty]{\,}& \left(\begin{array}{c}
E_R\,  e^{i k_1 x}\\
E_L\,  e^{i k_1 x}  
\end{array}\right), \label{right1}\\
\label{ER1}
E_R &\equiv& e_R  + \frac{4 d^{+} e^{i\theta_0}}{c_0},\\
E_L &\equiv& e_L +\frac{4 d^{-} e^{-i\theta_0}}{c_0}.
\label{EL1}
\er
As observed from the
expressions above  they are  reflectionless. In fact, we are assuming a plane wave incident from the left with components $e_{R,L}\, e^{i k_1 x}$  and vanishing coefficients of reflection $e'_{R,L} =0$. However, the transmission coefficients are non-vanishing, i.e. $(\frac{4 d^{\pm}}{c_0}) \neq 0$. The probability densities of these solutions at $x= \pm \infty$, respectively, are
\br
\left[\zeta^{\star}_R \zeta_R+\zeta^{\star}_L \zeta_L\right](x= -\infty) &=&  \Big[\zeta^{\star\,(free)}_R \zeta^{(free)}_R + \zeta^{\star\,(free)}_L\zeta^{(free)}_L\Big](x= -\infty) \label{lr10}\\
&=& e^\star_R  e_R +  e^\star_L  e_L,
\er
and
\br
\left[\zeta^{\star}_R \zeta_R+ \zeta^{\star}_L \zeta_L \right](x= + \infty) &=& \Big[\zeta^{\star\,(free)}_R \zeta^{(free)}_R + \zeta^{\star\,(free)}_L\zeta^{(free)}_L\Big](x= +\infty) \\
&=& e^\star_R  e_R +  e^\star_L  e_L +\nonumber\\
&&
 \frac{4}{c_0^2}\left[4 (d^{+\,\star} d^{+}+ d^{-\,\star} d^{-}) + c_0 e^{i \theta_0} (d^{+\,\star} e_R + e^\star_L d^{-})+  c_0 e^{-i \theta_0} (d^{-\,\star} e_L + e^\star_R d^{+})\right].\nonumber\\
\label{lr1}
\er    
Therefore, both quantities presented above must be equal, in accordance with the principle of probability conservation. Thus, we have
\br
\label{zero1}
4 (d^{+\,\star} d^{+}+ d^{-\,\star} d^{-}) + c_0 e^{i \theta_0} (d^{+\,\star} e_R + e^\star_L d^{-})+  c_0 e^{-i \theta_0} (d^{-\,\star} e_L + e^\star_R d^{+}) =0.
\er
Note that the relationship (\ref{zero1}), which follows from (\ref{lr10})-(\ref{lr1}), is crucial for matching both sides of equation (\ref{sta31}), as the derivative of the scalar field vanishes at $x= \pm\infty$.
 
Moreover, unitarity requires the coefficients to
satisfy
\br
\label{unit1}
 e^\star_R  e_R +  e^\star_L  e_L =1.
\er  

So, the scalar and spinors defined by the relationships (\ref{tau1f}) and (\ref{ztau1})-(\ref{ztau2}), respectively, together with the tau functions (\ref{stau0})-(\ref{staul}) satisfy the system of equations (\ref{sta11})-(\ref{sta31}) provided that 
\br
\label{pars11}
E_1 &=&\pm \sqrt{k_1^2+ M^2} ,\,\,\,\,\,\,\,\,\ k = - E_1\,\tan{\theta_0},\,\,\,\,\,\, e^\star_L = e_L\\
\label{pars12}
d^{+}  &=& \frac{1}{2} c_0 e_L  e^{i \theta_1}\frac{M \sin{\theta_0}}{k_1 +i E_1 \tan{\theta_0} } ,\,\,\,\,\, d^{-} = \frac{i}{2} c_0 e_L  \frac{ (k_1 - E_1)\sin{\theta_0}}{k_1 +i E_1 \tan{\theta_0} },\\
\label{pars13}
e_R &=& - i e^{i \theta_1} e_L \frac{k_1- E_1}{M}, \,\,\,\,\,\ e_L = \frac{\sqrt{k_1 + E_1}}{\sqrt{2 E_1}}\\
\label{pars14}
\l &=&  \b (\frac{\cos^2{\theta_o}}{2E_1}) \(\frac{M^2}{E_1^2 - M^2\cos^2\theta_o}\).
\er
The parameters above satisfy the relationships (\ref{zero1}) and  (\ref{unit1}) for an arbitrary value of the real  parameter $c_0$. 

In the Figs.  7 and 8 we plot the real components of the scattering states $\zeta_a,\,a=1,..,4\, (\zeta_R \equiv \zeta_3+i \zeta_4,\,\zeta_L \equiv \zeta_1+i \zeta_2)$, for $E_1 = \mp 1.01119, k_1=0.15, M=1,  \theta_o = \pi/4,\beta=1,\theta_1 = \pi/8$. Notice that each component of the incident free wave undergoes a distortion at the origin and a phase shift due to its interaction with the soliton. Despite this distortion and phase shift, the wave’s amplitude remains unchanged after the  transmission  through the soliton field. Moreover, the scattering process is completely reflectionless, meaning that no part of the wave is reflected. The wave passes through the soliton entirely, without any loss of energy or change in amplitude.
 
Interestingly, a notable property of linearized integrable systems around a soliton is their ability to support reflectionless scattering states \cite{koller}. Specifically, in the spinor sector of the integrable ATM model, the equations governing the system become linear with respect to the spinor fields. It can be argued that this linear system leads to a reflectionless scattering of the spinor field around the soliton, regardless of the energy involved. On this point, it is worth noting that the spinor sector of the ATM model closely resembles the Bogoliubov-de Gennes equation, in which the reflectionless property of self-consistent multi-soliton solutions has been established \cite{takahashi2}.

In summary, an analysis of our exact analytical results reveals that scattering in our soliton-fermion system is reflectionless. Additionally, narrow solitons induce stronger distortions and produce larger phase shifts in the scattering states (see Figs. 7 and 8), whereas broad solitons exert weaker effects on the scattering behavior.
 
\begin{figure}
\centering
\label{fig6k}
\includegraphics[width=1.5cm,scale=4, angle=0,height=3.8cm]{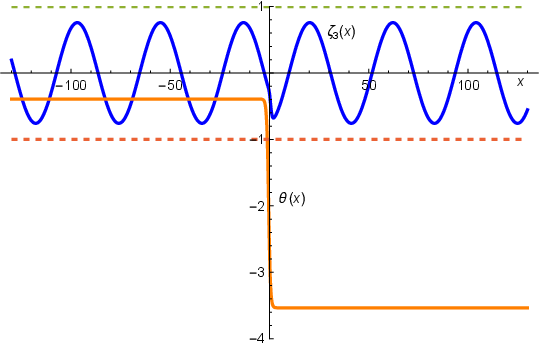} 
\includegraphics[width=1.5cm,scale=4, angle=0,height=3.8cm]{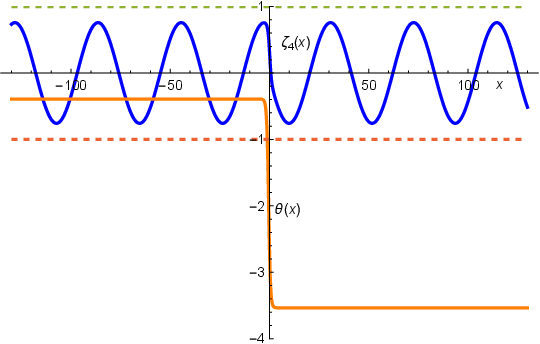}
\includegraphics[width=1.5cm,scale=4, angle=0,height=3.8cm]{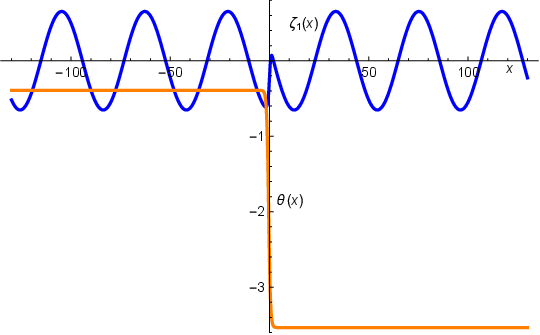}
\includegraphics[width=1.5cm,scale=4, angle=0,height=3.8cm]{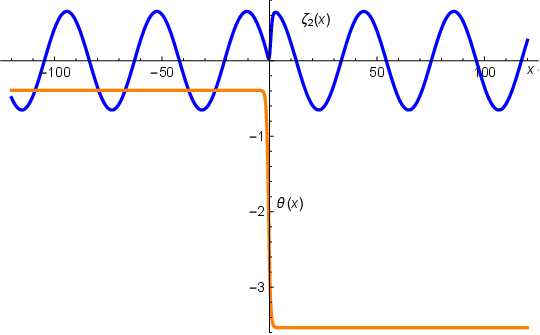}
\parbox{6in}{\caption{(color online) Components of the scattering states $\xi_a,\,a=1,..,4\, (\xi_R \equiv \xi_3+i \xi_4,\,\xi_L \equiv \xi_1+i \xi_2)$. For $E_1 = -1.01119, k_1=0.15,M=1, c_0=2.5, \theta_o = \pi/4,\beta=1,\theta_1 = \pi/8.$ Note that the waves undergo a phase shift due to presence of the soliton at the center.}}
\end{figure}

\begin{figure}
\centering
\label{fig7k}
\includegraphics[width=1.5cm,scale=4, angle=0,height=3.8cm]{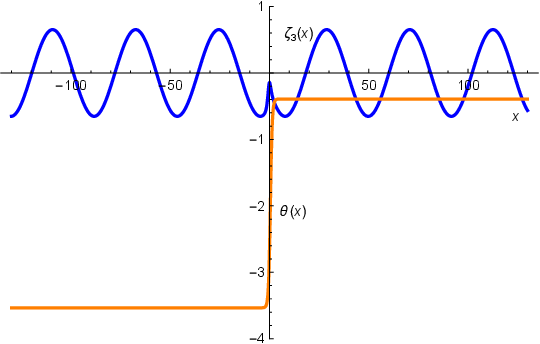} 
\includegraphics[width=1.5cm,scale=4, angle=0,height=3.8cm]{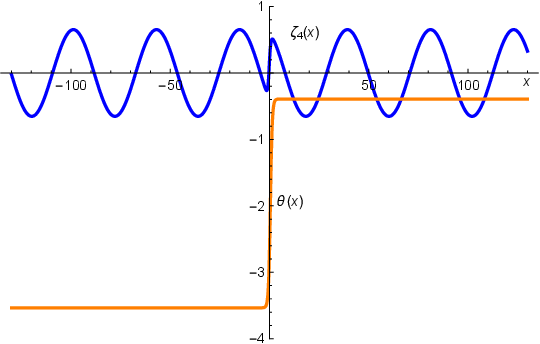}
\includegraphics[width=1.5cm,scale=4, angle=0,height=3.8cm]{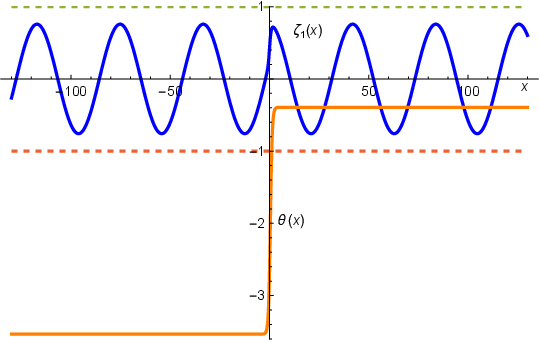}
\includegraphics[width=1.5cm,scale=4, angle=0,height=3.8cm]{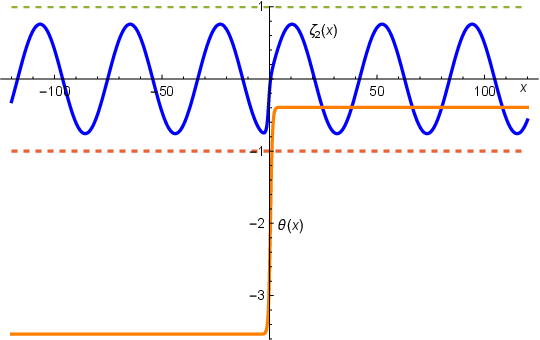}
\parbox{6in}{\caption{(color online) Components of the scattering states $\xi_a,\,a=1,..,4\, (\xi_R \equiv \xi_3+i \xi_4,\,\xi_L \equiv \xi_1+i \xi_2)$. For $E_1 = 1.01119, k_1=0.15,M=1, c_0=2.5, \theta_o = \pi/4,\beta=1,\theta_1 = \pi/8.$ Note that the waves undergo a phase shift due to presence of the soliton at the center.}}
\end{figure} 

\subsection{Phase shift and Levinson's theorem}

In order to compute the phase shift of the outgoing spinor scattering state with respect to the incoming spinor state one must identify these components entering into the solution (\ref{ztau1})-(\ref{ztau2}) for  the spinor field of the model at $x \rightarrow \pm \infty$ represented in (\ref{left1})-(\ref{EL1}). So, from the equations (\ref{left1})-(\ref{EL1}) and the parameter relationships (\ref{pars11})-(\ref{pars14}) one can write the incoming free spinor field at $x \rightarrow -\infty$ and a component of the outgoing spinor wave at $x \rightarrow +\infty$ carrying the effect of the soliton, respectively, as  
\br
\label{smat}
\left(\begin{array}{c}
e_R\, e^{i k_1 x}\\
e_L\, e^{i k_1 x} 
\end{array}\right) \er
and 
\br \label{trans0}
 \left(\begin{array}{c}
e_R\, e^{i k_1 x}\\
e_L\, e^{i k_1 x} 
\end{array}\right) &+&  \,\left(\begin{array}{c}
- 2 i\,  e^{i \theta_o} \, \frac{\sin{(\theta_o)} (k_1 + E_1)}{k_1+i E_1 \tan{\theta_o}} \,\, e_R\, e^{i k_1 x}\\
2 i \, e^{-i \theta_o} \, \frac{\sin{(\theta_o)} (k_1 - E_1)}{k_1+i E_1 \tan{\theta_o}}\,\,  e_L\,  e^{i k_1 x} \end{array}\right)=\\
\label{trans1}
 \left(\begin{array}{c}
e_R\, e^{i k_1 x}\\
e_L\, e^{i k_1 x} 
\end{array}\right) &-&  \, e^{i \d(k_1)} e^{i \theta_o \s_3}\left(\begin{array}{cc}
(E_1 + k_1) \widetilde{\D} & 0\\
0 & (E_1-k_1) \widetilde{\D}  
\end{array}\right)\,\left(\begin{array}{c}
e_R\, e^{i k_1 x}\\
e_L\,  e^{i k_1 x}
\end{array}\right),\er
with
\br
\s_3 = \(\begin{array}{cc} 1  & 0\\
0 & -1\end{array}\), \,\,\,\,\,\,\,\,\,\,\, 
\widetilde{\D} &\equiv & \frac{2 \sin{\theta_o}}{\sqrt{k_1^2+ E_1^2 \tan^2{\theta_o}}}, \label{s33}\er
and $\d(k_1)$ given by
\br
\label{delta11}
\d(k_1) &\equiv& \arctan{\Big[ \frac{k_1}{E_1 \tan{\theta_o}}\Big]},\,\,\,\,\,\, E_1^2 = k_1^2 + M^2.
\er
Note that when $\theta_o =0$ ($\widetilde{\D} = 0$) is considered in the components (\ref{trans1}), meaning there is no soliton background, the spinor components carrying the effect of the soliton vanish. Consequently, the solution (\ref{ztau1})–(\ref{ztau2}) reduces to the free spinor field continuous background. In fact, the soliton tau functions $\tau_{R,L}$ in (\ref{staur})-(\ref{staul}) vanish in this limit since one has $d^{\pm}=0$ in (\ref{pars12}) for $\theta_o =0$.  
 
Notably, the factor $\widetilde{\D}$ in the scattering S-matrix in (\ref{trans1}), when setting $k_1 \equiv i k$, exhibits singularities at $E_1 = - k \cot{\theta_o}$. So, one can argue that transmission coefficients in (\ref{trans0}) have simple poles at the imaginary momenta of the bound state energies. These singularities correspond to the spinor bound states and the soliton constructed in (\ref{epsi11})-(\ref{mth11}). Thus, our results are consistent with Levinson’s theorem, which asserts that each bound state emerges from a continuum state of the unperturbed (free) system.

Comparing the expressions (\ref{smat}) and (\ref{trans1}) one observes that the spinor components develop different phase shifts, i.e. 
\br
\label{phs1}
\d_1 &=& \d(k_1) + \theta_o ,\\
\d_2 &=& \d(k_1) - \theta_o,\label{phs2}
\er
for the upper and lower components, respectively. The challenge in defining phase shifts becomes evident when the background field has arbitrary boundary values: under the conventional definition, the upper and lower components generally acquire different phase shifts.
 
It is known in the literature that a prescription must be provided in order to compute an unique phase shift. We follow the prescription proposed in \cite{Gousheh1} in which the phase shift is defined as an average of the two quantities in (\ref{phs2})-(\ref{phs2}). In fact, in \cite{Gousheh1} a version of the ATM model without the kinetic term of the scalar field has been studied, assuming a prescribed scalar field soliton. Then, one has a phase shift defined as
\br
\d(k_1) &=& \frac{1}{2}\(\d_1(k_1) + \d_2(k_1)\).
\er    
   
Then, from (\ref{delta11}) one has
\br
\label{levin0}
 \d(\infty) - \d(0) = \frac{(2 n + 1) \pi}{2} - \theta_o,\, \,\,\, n\in \IZ.
\er
This result bears similarity to that presented in \cite{Gousheh1}; however, our formulation provides a complete treatment of the problem, as both the spinor and scalar fields are treated as dynamical variables. Notably, our Lagrangian includes the scalar field’s kinetic term, and the coupled system has been solved in a fully self-consistent manner using the tau function approach to deal with scattering sates.

\subsection{Fermion vacuum polarization energy (VPE) and the scattering states}

The fermion vacuum polarization energy  arises from modifications to the fermionic energy spectrum induced by the presence of the soliton. In systems like the ATM model, where the soliton induces reflectionless scattering (a common feature in many integrable models), the fermionic phase shifts in the scattering spectrum are simplified. This simplification makes the calculation of the vacuum polarization energy more manageable, as we will demonstrate below. Such systems often permit exact or semi-analytical solutions. 

The vacuum polarization energy, encompassing contributions from an infinite number of modes, requires careful regularization and renormalization to eliminate divergences. Through the renormalization process, only the finite, physically meaningful energy contributions are retained. The vacuum polarization energy plays a crucial role in shaping the stability and dynamics of soliton-fermion systems. By typically reducing the system's total energy, it enhances the soliton's stability through its interaction with the quantum fluctuations of the fermion field. 

The vacuum polarization energy (VPE) for the spinor sector of the ATM model (\ref{atm0}) was previously computed for a static and {\sl prescribed} piecewise linear pseudo-scalar background field \cite{mohammadi}. Using the exact fermionic spectrum for this setup, the VPE was obtained by subtracting the vacuum energy with and without the background field. Additionally, the spinor sector of the ATM model with a {\sl prescribed} sine-Gordon type soliton as the background field was analyzed through numerical simulations and the phase shift method to determine the total Casimir energy \cite{mohammadi1}. In the present work, we focus on exact solutions that incorporate the back-reaction of the spinor field on the true soliton of the model.

In the exactly solvable soliton-fermion system considered here, all normalized continuum wave functions with negative energy in the presence of the soliton, $\theta(x)$, have been explicitly calculated above. The vacuum polarization energy (VPE) can then be determined exactly and directly by subtracting the vacuum energy of the system without the soliton from that with the soliton, where the soliton acts as the disturbance. To analyze this in detail, let us consider  \cite{mohammadi, mohammadi1}  
\br
\nonumber
<\Omega|H|\Omega> - <0|H_{free}|0> &=& \int_{-\infty}^{+\infty} dx  \int_{0}^{+\infty} \frac{dp}{2\pi} (-\sqrt{p^2 + M^2}) \, \zeta^{\star}_{p}  \zeta_p -  \\
&&
\int_{-\infty}^{+\infty} dx  \int_{0}^{+\infty} \frac{dk_1}{2\pi} (-\sqrt{k_1^2 + M^2}) \,  \zeta^{\star\,(free)}_{k_1}  \zeta^{(free)}_{k_1} \label{vpe1} \\
&=&  \int_{0}^{+\infty} dk_1 (-\sqrt{k_1^2 + M^2}) [\hat{\rho}^{(sea)}(k_1) - \hat{\rho}^{(sea)}_{0}(k_1)].\label{vped}
\er
The functions $\zeta_p$ and $\zeta^{(free)}_{k_1}$ stand for normalized wave functions for the continuum states with negative energy in the presence and absence of the soliton, respectively. The factor $ [\hat{\rho}^{(sea)}(k_1) - \hat{\rho}^{(sea)}_{0}(k_1)]$ in the eq. (\ref{vped}) measures the spectral deficiency in the continuum states and it is the difference between the
density of the continuum states with the negative energy in
the presence and absence of the kink.  

The divergent integrals above have been formally manipulated, and the prescription for subtracting the two divergent integrals of  (\ref{vpe1}) is to subtract the integrands with corresponding values of $p = k_1$ and then evaluate
the leftover $x-$integral, providing the finite result in (\ref{vped}).

Let us mention that indirect approaches, such as the phase shift method, are sometimes employed to calculate the VPE in (\ref{vped}). This method links the derivative of the phase shift with respect to momentum to the spectral deficiency in the continuum states. Below we resort to the phase shift approach in order to compute the VPE in (\ref{vped}). So, one has 
\br
\label{levinson1}
\frac{1}{\pi} \frac{d}{dk_1}\d^{sea}(k_1) =
\hat{\rho}^{(sea)}(k_1) - \hat{\rho}^{(sea)}_{0}(k_1)\er 

Next, taking into account  (\ref{levinson1}) the VPE equation (\ref{vped}) can be written as  
\br
\label{vped0}
VPE &=&<\Omega|H|\Omega> - <0|H_{free}|0> \\
&=& \int_{0}^{+\infty} \frac{dk_1}{\pi} (-\sqrt{k_1^2 + M^2}) \, \frac{d\d^{sea}(k_1)}{dk_1}  \label{vped1}\\
&=& \int_{0}^{+\infty} \frac{dk_1}{\pi} (-\sqrt{k_1^2 + M^2}) \, \frac{d}{dk_1} (\d^{sea}(k_1)-\d^{sea}(+\infty)) \label{vped11}\\
&=&  \int_{0}^{+\infty} \frac{dk_1}{\pi} \frac{k_1}{\sqrt{k_1^2 + M^2}} \, (\d^{sea}(k_1)-\d^{sea}(+\infty)) + \frac{M}{\pi} (\d^{sea}(0)-\d^{sea}(+\infty))\label{vped2}
\er
SO, taking into account $\d^{sea}(k_1)$ in (\ref{delta11}) for $E_1= - \sqrt{k_1^2 + M^2}$, the above integration furnishes
\br
\label{vpef0}
VPE &=& \frac{1}{2} M \cos{\theta_o} + \frac{M}{\pi} (\d^{sea}(0)-\d^{sea}(+\infty)) \\
&=&  \frac{1}{2} M \( \cos{\theta_o} +\frac{2 \theta_o}{\pi}  \) - \frac{1}{2} M. \label{vpef2}
\er
In the last line it has been used the expression (\ref{levin0}) with $n=0$. So, this is the VPE in the presence of the kink (antikink) with topological charges $+\frac{2 \theta_o}{\pi}$\, ( $- \frac{2 \theta_o}{\pi} $).  Observe that when $\theta_o =0$ (no bound states present), corresponding to the free case, the vacuum polarization energy (VPE) vanishes, i.e., $VPE=0$. Consequently, the last term in (\ref{vpef2}), which accounts for the threshold states at $E_1= \pm M$, is crucial to ensure the VPE correctly vanishes in the fermion-free case when $\theta_o =0$. 

Notably, the $k_1$ dependence of the integrand in (\ref{vped1}) resembles that of equation (3.8) in \cite{mohammadi}, observed in their  model under a specific limit. Specifically, this similarity arises in the special slope limit $\mu \rightarrow \infty$ for the piecewise linear scalar field in the central region. In order to examine this similarity, let us substitute the phase shift (\ref{delta11}) into (\ref{vped1}) to get 
\br
 \int_{0}^{+\infty} \frac{dk_1}{\pi} (-\sqrt{k_1^2 + M^2}) \, \frac{d\d^{sea}(k_1)}{dk_1} = \int_{0}^{+\infty} \frac{dk_1}{\pi} \frac{M^2 \sin{\theta_o} \cos{\theta_o}}{k_1^2 + M^2 \sin^2{\theta_o}}.\label{compare1}
\er
The integration in  the r.h.s. of (\ref{compare1}) is the same as in eq. (3.8) of reference \cite{mohammadi} up to a constant value. However, since the piecewise linear field configuration chosen in \cite{mohammadi} is not a true soliton solution to the field equations, our exact expression in (\ref{vpef2}) fully incorporates the characteristics and effects of a genuine soliton.

In the Fig. 9 we present the plot of the vacuum polarization energy VPE  (\ref{vpef2})  as  $VPE(\theta_o)\, \, vs \,\, \theta_o$. Comparing this figure with the corresponding Figures 5 and 6, which depict the energy $E = E_{kf} + \epsilon $ (red lines) and the bound-state energy $\epsilon$ (magenta lines), reveals that the contribution of the vacuum polarization energy (VPE) must be given equal consideration alongside $E_{kf}$ and $\epsilon$. This is because, for certain values of the parameters $\theta_o \sim \b^2 $ (see (\ref{thetao2})), the VPE is of the same order of magnitude as $E_{kf}$ and $\epsilon$ energy components. The total energy will be analyzed in detail below.

\begin{figure}
\centering
\label{fig8k}
\includegraphics[width=1.5cm,scale=4, angle=0,height=4.5cm]{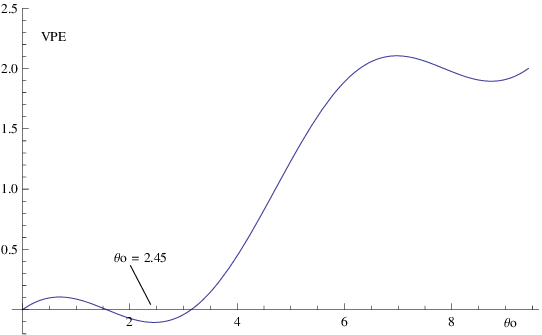}     
\parbox{6in}{\caption{(color online) Plot of the VPE in (\ref{vpef2}) for $M=1$. The figure shows $VPE(\b)$ vs $\theta_o$. Note the appearance of the minimum at $\theta_o= 2.45$.}}
\end{figure} 
Some comments are in order here regarding the above calculation of the VPE as compared to the one in Refs. \cite{farhi, farhi2}. First, the integrand in (\ref{vped1}), once the phase shift (\ref{delta11}) is used, takes the form of the r.h.s. of eq. (\ref{compare1}). This integral is finite, since its integrand at $k_1 \rightarrow +\infty$ behaves as $\frac{1}{k_1^2}$. Therefore, in our approach this contribution to the VPE is finite. Second, note that our model becomes a sub-model of the one in \cite{farhi, farhi2} provided that their scalar fields $\phi_{1,2}$ lie on the chiral circle $(\phi_1\,,\,\phi_2) = \frac{1}{2 \hat{\b}} (\cos{2 \hat{\b} \vp}\, ,\, \sin{2 \hat{\b} \vp})$, where $\vp(x \rightarrow -\infty) \rightarrow 0$ and $\vp(x \rightarrow +\infty) \rightarrow \pi/\hat{\b}$, with $\vp$ being the ATM scalar in (\ref{atm0}). Third, in \cite{farhi, farhi2} the fermion effective energy in the presence of the classical background has been computed numerically in the ﬁeld theory approach. Standard perturbative renormalization procedure has been performed through one loop order to the VPE (\ref{vped0})-(\ref{vped1}). Fourth, their entire counterterm contribution to the phase 
shift becomes $\hat{\d}(k_1) \equiv \frac{8M^2}{k_1} \int_{0}^{\infty}\, dx  (\vec{\phi}(x)^2-\frac{1}{4 \hat{\b}^2})$. On the chiral circle $\vec{\phi}^2 = \frac{1}{4 \hat{\b}^2}$, connected to the results presented here, this counterterm contribution to the VPE vanishes, implying that the one-loop quantum contribution to the energy is finite. Fifth, for scalar configurations on the chiral circle and through numerical computation, it has been observed in \cite{farhi2} that $\d(k_1)$ goes like $\frac{1}{k_1^3}$ for $k_1$ large. This would imply the integrand in the l.h.s. of (\ref{compare1}) to decrease more rapidly than $\frac{1}{k_1^2}$. By contrast, our exact analytical result shows that the integrand decreases as $\frac{1}{k_1^2}$ for $k_1$ large. Sixth,  one can argue that our analytical result in (\ref{vpef2}) for the VPE is an exact result at the one-loop order in the field theory approach of Refs. \cite{farhi, farhi2}.    

\subsection{Total energy and stability of the solutions}

In this sub-section we consider the energy contributions of the fermion-kink configuration, the valence
fermion and explore the effect of the VPE energy on the total energy. 

We emphasize that the model under investigation admits,  simultaneously, a classical scalar soliton and localized fermionic bound states, resulting from a chiral coupling between the fermionic and scalar fields. The topological charge is dynamically generated and it depends on the coupling constant. Crucially, the scalar sector lacks a self-interaction potential ${\large -}$ unlike the model studied in \cite{farhi,farhi2} with self-coupling potential and no classical solitons. The authors study soliton formation only through quantum stabilization mechanisms. Our model is the truncation of \cite{farhi, farhi2} with vanishing scalar self-coupling, such that our classical solitons belong to the full theory in suitable parameter space.

However, upon incorporating quantum effects, the classical treatment must be revisited. In particular, the spatially varying soliton configuration, together with the fermion bound state, should be regarded as minimizing an effective energy that includes both classical contributions and quantum corrections arising from vacuum fluctuations. 

Next, our objective is to find the points of absolute and relative minima of the effective energy for some values of the ATM coupling constant $\hat{\b}$ in (\ref{atm0}).

So, the total energy consists of three components: the classical fermion-soliton interaction energy $E_{kf}$, the energy of the bound-state fermion $\epsilon$, and the fermion vacuum polarization energy VPE. Then, the VPE (\ref{vpef2})  must be added to the energy $E$ in (\ref{E2})  in order to compute the total energy $E_{tot}$ as follows
\br
\label{Etot1}
E_{tot} &=& E + VPE\\
\label{Etot2}
&=& E_{kf} + \epsilon + VPE\\
\nonumber
&=&  -M \,\frac{(2 \a -\b)(2\a + 7\b)}{\b^2 (2 \a +\b)^2} \,\big[2 \, \theta_o \cos{\theta_o} + \sin{\theta_o} -  \sin{(3 \theta_o)}\big] + M \cos{\theta_o} + \\
&& \frac{1}{2} M \( \cos{\theta_o} +\frac{2 \theta_o}{\pi}  \) - \frac{1}{2} M,\label{Etot3}
\er
with 
\br  
\a = (r_{\pm})\b,\,\,\,\,\, r_{\pm} =-5/2 \pm \sqrt{7},\,\,\,\,\, \b = \pm  \sqrt{\frac{8}{2 r_{\pm}+1}}\, \sqrt{\theta_o}.
\label{Etot33}
\er
In the Fig. 10 we plot  the total energy $E_{tot}$ (\ref{Etot3}) for the kink-fermion system in terms of $\theta_o$  ( $E_{tot} \,vs \, \theta_o$)  for the both cases $\a = r_{\pm} \b$, blue line and dashed line, respectively. Observe that there is only a very slight difference between the two figures. By analyzing the total energy depicted in these graphs, we can investigate the system's stability. The figures reveal an absolute minima at $\theta_o \approx 2.9$, indicating that these configurations are not only energetically favorable but also stable against small fluctuations of $\theta_o$. Relative minima occur at approximately the same points $\theta_o \approx 9.3, 15.6,$ for the both figures.  

It is instructive to see the dependence  on $\theta_o$ of the initial ATM Lagrangian coupling constant $\hat{\b}$ in (\ref{atm0}). From (\ref{lab11}) and (\ref{thetao2}) the coupling parameter $\hat{\b}$ in terms of $\theta_o$ can be written as  
\br
\nonumber
\hat{\b} (\theta_o)  &=&  -\frac{e_1}{3+ e_2 \sqrt{7}} \b \\
&=& \mp \frac{2 e_1\sqrt{2}}{(3+ e_2 \sqrt{7})(\sqrt{2 r_{\pm} + 1})}  \sqrt{\theta_o}, \,\,\,e_a = \pm 1,\,\,a=1,2. \label{betac}
\er 
This yields a total of sixteen distinct values of the ATM coupling constant $\hat{\b}$ for the given set of parameters $\{r_{\pm}, \theta_o\}$.  
\begin{figure}
\centering
\label{fig9k}
\includegraphics[width=1.5cm,scale=4, angle=0,height=4.5cm]{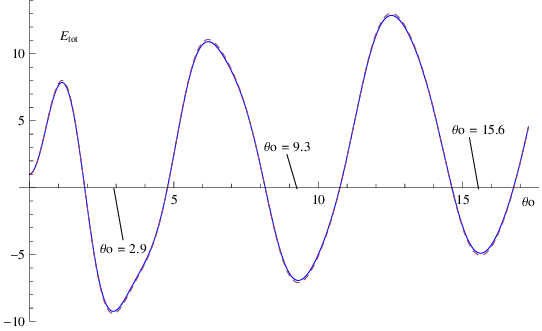}  
\parbox{6in}{\caption{(color online) The function $E_{tot}(\theta_o)$ of the total energy (\ref{Etot3}) plotted for $M=1$. The blue line stands for $r_{+}$ and the dashed red line for $r_{-}$. Observe that the both figures display an absolute minima at $\theta_o \approx 2.9$ and relative minima at approximately the same points $\theta_o \approx 9.3, 15.6,$ for the both figures.}}
\end{figure}

\section{Discussions and conclusions}
\label{sec:discuss}

The $sl(2)$ affine Toda model coupled to matter (ATM) (\ref{atm0}) presents a rich framework for studying the interplay between bosonic and fermionic fields, especially within the context of integrable systems. Through the Faddeev-Jackiw symplectic Hamiltonian reduction, we have elucidated the complex dynamics governing this model, notably the intricate relationship between constraints, symplectic potentials, nonlinearity, topology and the strong-weak dual coupling sectors. This work emphasizes the significance of ensuring the equivalence between Noether and topological currents, a key issue in understanding the model's underlying symmetries and conservation laws.

One of the findings of this study is the emergence of fermion excited bound states localized on the kinks of the reduced ATM model (\ref{thetaeq1100ra})-(\ref{bpsieq1ira}). The bound states with charge densities (\ref{mth00}) and (\ref{mth11}) are not merely mathematical artifacts; they play an active role in the system's dynamics by contributing to back-reaction effects through the redefined constants $\{\l, \a, \b\}$, which according to (\ref{lab1})-(\ref{lab11})) depend on the coupling constant $\hat{\b}$. This back-reaction significantly alters the topological properties of the model, introducing a novel pumping mechanism for the topological charge of the kink. This mechanism, driven by the fermionic back-reaction, highlights the dynamic interplay between fermionic excitations and topological features, offering new insights into the non-trivial topology of integrable models and their deformations, as the emergence of the non-integrable DSG model (\ref{pot11}) in the scalar decoupled regime of the  reduced ATM model. So, the topological charge pumping mechanism represents, to our knowledge,  a novel advancement in the study of non-linear dynamics and topological effects in $1+1$ field theories.

Our analysis also highlights the power of tau function techniques in constructing self-consistent solutions within the ATM model. These techniques provide a robust framework for exploring how the properties of kinks, fermionic bound states and scattering states depend on various model parameters. Such insights are invaluable for understanding the stability and behavior of solitonic solutions in integrable systems, which are often characterized by their sensitivity to changes in parameters and external conditions.

Moreover, the study shows  that the inclusion of new parametrizations for scalar and Grassmannian fermionic fields leads to a deeper understanding of the fermion-scalar model's physical implications. By examining the model, we have uncovered the essential roles these fields play in shaping the model's dynamics and topological characteristics. So, our work contributes to a broader understanding of how modifications in field parametrizations and the fermion bound states excitations can impact the behavior and properties of integrable systems and of their non-integrable modifications.

Our exploration of the ATM  model has unveiled a wealth of phenomena that deepen our understanding of kink-fermion systems. Our approach differs from the Bogomolnyi trick, which gets first-order equations by completing the square in the energy functional. Instead, our method parallels the approach proposed in \cite{devega}, where the first-order equations for vortices in $1+2$ dimensions arise provided that  the conservation of the energy-momentum tensor is assumed. Our findings show  the importance of considering back-reaction effects and their influence on the appearance of the in-gap fermion-kink $E_{kf}$ energy (\ref{E2})-(\ref{E2i}), fermion bound state energy $\epsilon$ (\ref{E2i}) and the energy $E_{DSGkink}$ of the decoupled scalar kink states in the continuum (KIC) (\ref{E31}) for some regions in parameter space. 

We have added the fermion vacuum polarization energy (VPE) (\ref{vpef2}) to the energy of the soliton-fermiom system, such that the total energy (\ref{Etot3}) comprises the classical fermion-soliton interaction energy $E_{kf}$, the energy of the bound-state fermion $\epsilon$, and the fermion VPE. We have concluded that the contribution of the VPE energy is not negligible in comparison to the valence fermion energy and it must be given equal consideration alongside $E_{kf}$ and $\epsilon$. Moreover, examining the total energy we have explored the stability points of the system under small variations of the coupling constant $\hat{\b}$. In Fig. 10, the stability points are indicated as the global minimum and local minima of the total energy plot as $E_{tot}\,vs\, \theta_o$.
 
Several avenues for future research remain. These may include extension to ATM models based on higher order affine Lie algebras, which may reveal additional structure and symmetries. Quantum corrections and fermion vacuum polarization effects would be an intriguing direction for future research in these models. It would be interesting to analyze the emergent non-integrable models driven by the fermionic back-reaction, in the context of the quasi-integrability concept \cite{ferreiraq, np20} and performing numerical simulations to verify and extend the analytical results, particularly in regimes where analytical solutions are challenging to obtain. 

Moreover, exploring potential experimental systems that can replicate the behavior of the ATM model-such as those found in condensed matter or optical settings-would be valuable for validating its theoretical predictions. It is also important to investigate how the topological charge pumping mechanism might influence other domains, including quantum computing and condensed matter physics, where topological states play a central role. The rich interplay between topology, non-linear dynamics, and fermionic excitations remains a promising avenue for discovery, with the potential to unveil novel insights and applications in the future.

\[\]
\noindent {\bf Acknowledgements}

We thank the FC-UNI (Lima-Per\'u) for hospitality during the first stage of the work. RQ thanks Concytec (Per\'u) for financial support and  professor R. Metzger (IMCA-UNI) for making possible his visit to the IF-UFMT (Cuiab\'a-Brazil).

{\bf Declarations}
 
{\bf Data Availability} This manuscript has no associated data
or the data will not be deposited. [Authors’ comment: The data that
support the findings of this study are available from the corresponding
author, upon reasonable  request.]

{\bf Code Availability Statement} The manuscript has no associated
code/software. [Author’s comment: Code/Software sharing not applicable to this article as no code/software was generated or analyzed during
the current study.]

{\bf Conflicts of Interest}: The authors declare no conflict of interest.

\appendix

\section{The Faddeev-Jackiw formalism}
\label{app:FJ}

The Faddeev-Jackiw (F-J) approach \cite{fj} and symplectic methods \cite{barc1} offer a direct way to handle constraint systems without requiring the classification of constraints into first and second class. Below is a brief overview of the F-J  method. We begin with a first-order Lagrangian in time derivatives, which may originate from a usual second-order Lagrangian by introducing auxiliary fields. The general form of such a Lagrangian is 
\br
\lab{lag1}
L=a_{i}(\xi)\dot{\xi}^{i}-V(\xi).
\er
Where the coordinates $\xi_{i}$, with $i=1,...,N$, stand for the generalized coordinates. Notice that when a Hamiltonian is
defined by the usual Legendre transformations, V may be identified with the Hamiltonian H. 
   
The first order system (\ref{lag1}) is characterized by a closed two-form. If the two-form is not degenerated, it defines a symplectic
structure on the phase space M, described by the coordinates $\xi_{i}$. On the other hand, if the two-form is singular, with
constant rank on M, it is called a (pre)symplectic two-form. Thus, in terms of components, the (pre)symplectic form
is defined by
\br
\lab{form}
f_{ij}(\xi) \equiv \frac{\pa}{\pa{\xi}^{i}}a_{j}(\xi)-\frac{\pa}{\pa{\xi}^{j}}a_{i}(\xi),
\er 
with the vector potential $a_{i}(\xi)$ being an arbitrary function of $\xi^{i}$. The Euler-Lagrange equations are given by   
\be
\lab{eqm1}
f_{ij}\dot{\xi}^{j}\equiv \frac{\pa}{\pa{{\xi}}^{i}}V({\xi}). 
\ee
In the non-singular, unconstrained case the anti-symmetric $N{\mbox x}N$ matrix $f_{ij}$ has the matrix inverse $f^{ij}$, 
then $N=2n$, and \rf{eqm1} implies
\be
\lab{eqm2}
\dot{\xi}^{i}\equiv f^{ij}\frac{\pa}{\pa{{\xi}}^{j}}V({\xi}),
\ee
and the bracket will be defined by 
\be
\{\xi^{i}\;,\; \xi^{j}\}\equiv f^{ij}.
\ee
In the case that the Lagrangian \rf{lag1} describes a constrained system, the matrix $f^{ij}$ is singular which means that
there is a set of relations between the velocities reducing the degrees of freedom of the system. Let us suppose that the
rank of $f$ is 2n, so there exist $N-2n=N^{\prime}$ zero modes ${\bf v}^{\a}$, $\a\equiv 1,...,N^{\prime}$. The system is then
constrained by $N^{\prime}$ equations in which no time-derivatives appear. Then there will be constraints that reduce the 
number of degres of freedom of the theory. Multiplying \rf{eqm1} by the (left) zero-modes ${\bf v}^{\a}_{i}$ of $f_{ij}$ we get   
\be
{\bf v}_{i}^{\a}f_{ij}\dot{\xi}_{j}\equiv {\bf v}_{i}^{\a}\frac{\pa V(\xi)}{\pa {\xi}_{i}}\equiv 0.
\ee
These (symplectic) constraints appear as algebraic relations
\be
\lab{const}
\O_{\a}\equiv{\bf v}_{i}^{\a}\frac{\pa V(\xi)}{\pa {\xi}_{i}}\equiv 0.
\ee
By using Darboux's theorem  one can show that an arbitrary vector potential, $a_{i}$, whose associated field 
strength $f_{ij}$ is non-singular, can be mapped by a coordinate transformation onto a potential of the form
$a_{i}(\xi)\equiv \frac{1}{2}\xi^{j}w_{ji}$ with $w_{ji}$ a constant and non-singular matrix. Then, the Darboux construction may still be carried out for the non-singular projection of $f_{ij}$ given in \rf{form}. Then the Lagrangian becomes
\be
\lab{lagz}
L\equiv \frac{1}{2}\xi^{i}w_{ij}\dot{\xi}^{j}-V(\xi,z),
\ee
where $z$ denote the $N-2n$ coordinates that are left unchanged. Some of the $z^{\prime}s$ may appear non-linearly and
some linearly in \rf{lagz}. Then using the Euler-Lagrange equation for these coordinates we can solve for as many
$z^{\prime}s$ as possible in term of ${\xi}^{\prime}$s and other $z^{\prime}s$ and replace back in $V(\xi,z)$ so finally 
we are left only with linearly occuring $z^{\prime}s$. So, we can write the Lagrangian in the form   
\be
\lab{lagl}
L=\frac{1}{2}\xi^{i}w_{ij}\dot{\xi}^{j}-V(\xi)-\l_{k}\Phi^{k}(\xi),
\ee
where we have renamed the linearly occuring $z's$ as $\l_{k}$. We see that these $\l_{k}$ become the Lagrange multipliers
and $\Phi^{k}(\xi)$ are the constraints. To incorporate the constraints we solve the equations 
\be
\Phi^{k}(\xi)\equiv 0,
\ee
and replace back in \rf{lagl}. This procedure reduce the number of $\xi^{\prime}$s and we end up with a Lagrangian which
has the structure given in \rf{lag1}. Then the whole procedure can be repeated again until all constraints are eliminated
and we are left with a completely reduced, unconstrained and canonical system.

\section{First order equations imply the second order equation for $\theta$}
\label{app1}

Let us consider first the eq.  (\ref{psieq1}) and multiply it successively on the left by $\gamma_5$ and $\bar{\xi}$. Then, one gets
\br
\label{xi1}
\bar{\xi} \gamma^{\mu} \gamma_5  \pa_{\mu}\xi - i  M \bar{\xi}\gamma_5 \xi \cos{(\beta \theta)} -  M \bar{\xi}\xi \sin{(\beta \theta)} - 2g \bar{\xi} \gamma^{\mu} \g_5 \xi\Big[ (\frac{\alpha}{2g} - \l)\epsilon_{\mu \nu} \pa^{\nu} \theta - \l   \pa_{\mu} \hat{\S}\Big] =0.
\er
Similarly, consider (\ref{bpsieq1}) and multiply it succesively on the right by $\gamma_5$ and $\xi$. So one gets
\br
\label{xi2}
\pa_{\mu}\bar{\xi} \gamma^{\mu} \gamma_5  \xi - i  M \bar{\xi}\gamma_5 \xi \cos{(\beta \theta)} -  M \bar{\xi}\xi \sin{(\beta \theta)} +2 g \bar{\xi} \gamma^{\mu} \g_5 \xi \Big[ (\frac{\alpha}{2g} - \l)\epsilon_{\mu \nu} \pa^{\nu} \theta - \l   \pa_{\mu} \hat{\S}\Big] =0.
\er
Then, adding the both eqs.  (\ref{xi1}) and (\ref{xi2}), and multiplying by an overall factor $\frac{\b}{2}$, one can get
\br
\label{spinorf}
\frac{\b}{2} \pa_{\mu}\( \bar{\xi} \gamma^{\mu} \gamma_5 \xi \) - i \b M \bar{\xi}\gamma_5 \xi \cos{(\beta \theta)} - \b M \bar{\xi}\xi \sin{(\beta \theta)}=0.
\er   
In addition, one can write the equation (\ref{sigma1}) as
\br
\label{firstorder1}
 \epsilon^{\mu\nu}\pa_\nu \hat{\S} \equiv - \pa^{\mu} \theta + (\alpha+\frac{\beta}{2}-\iota)\, \bar{\xi}\gamma_5\gamma^{\mu}  \xi. 
\er 
Next, taking the derivative of the first order equation (\ref{firstorder1}) one can write the identity
\br
\label{chiral111}
\frac{\b}{2} \pa_{\mu}\( \bar{\xi} \gamma^{\mu} \gamma_5 \xi \) &=& \pa^2 \theta - (\a-\iota) \epsilon^{\mu\nu}\pa_{\mu} j_\nu,\\
&=& (\frac{1}{\a+ \frac{\b}{2}-\iota}) \pa^2 \theta, \label{chiral111i}
\er 
where we have used the identities $j_5^{\mu} = \epsilon^{\mu \nu} j_\nu\,\, (j_5^{\mu}  \equiv \bar{\xi} \gamma^{\mu} \gamma_5 \xi)$\, and $ \{\g_5 , \g^{\mu}\}=0$, as well as the expression in (\ref{sigma1}) for $j_\nu$. Note that the $\hat{\S}$ field does not appear in (\ref{chiral111i}) due to the identity $\epsilon^{\mu \nu} \pa_{\mu \nu} \hat{\S} =0.$ Then, taking into account  the expression  (\ref{chiral111i}) into  (\ref{spinorf}) one gets the second order differential equation
\br
\label{thetaeq1ap}
\pa^2 \theta  - 2 (\a+\b/2-\iota)\, M   \bar{\xi}\xi \sin{(\beta \theta)} -  2 i (\a+\b/2-\iota)\, M  \bar{\xi}\gamma_5\xi \cos{(\beta \theta)} &=&0.\er
This last equation (\ref{thetaeq1ap}) becomes identical to the equation of motion (\ref{thetaeq1}) for the scalar field $\theta $ provided that $\iota \equiv 0$. Thus, the set of first order equations (\ref{psieq1})-(\ref{bpsieq1}) and (\ref{sigma1}) (for the scalar field $\hat{\S} \equiv \S$) imply the second order differential equation (\ref{thetaeq1}) in the particular case $\iota =0$. One can argue that  for non-vanishing values of this parameter, i.e. $\iota \neq 0$ and free field $\hat{\S}$, one can not reproduce the second order differential  equation of motion (\ref{thetaeq1}) starting from the set of first order equations  (\ref{sigma1}) and (\ref{psieq1})-(\ref{bpsieq1}).

\end{document}